\input epsf
\documentclass[aps,prb,preprint,groupedaddress,floatfix]{revtex4}

\def\tens#1{\underline{\underline{#1}}}
\begin{document}

\title{Photoelasticity of crystalline and amorphous silica from first principles}
\author{D. Donadio, M. Bernasconi}
\affiliation{Dipartimento di Scienza dei Materiali and Istituto Nazionale per la  Fisica
dela Materia,  Universit\`a di  Milano-Bicocca, Via Cozzi 53, I-20125, Milano, Italy}
\author{F. Tassone}
\affiliation{Pirelli Cavi e Sistemi S.p.a., Viale Sarca 222, I-20126, Milano, Italy}

\begin{abstract}
Based on density-functional perturbation theory we have computed from first
principles the photoelastic
tensor of few crystalline phases of silica at normal conditions and high
pressure (quartz, $\alpha$-cristobalite,
$\beta$-cristobalite) and of models 
of amorphous silica (containig up  to 162 atoms), obtained by quenching from the melt in combined
classical and Car-Parrinello molecular dynamics simulations.
The computational framework has also
been checked  on the photoelastic tensor of crystalline silicon and MgO as prototypes of covalent
and ionic systems. The agreement with available experimental data is good.
 A phenomenological model suitable to describe the
 photoelastic properties of different silica polymorphs is devised by fitting on
 the ab-initio data.
\end{abstract}
\maketitle

\section{Introduction}

The dependence of the refractive index versus strain in silica glass, namely the photoelasticity,
is of interest from the fundamental point of view as well as for several technological applications in
optics and microelectronics.
For instance, photoelasticity in a-SiO$_2$ is known to cause a reduction of
fiber Bragg gratings efficiency \cite{limberger96} and to produce a loss of resolution
of pure silica lenses used in photolytography \cite{schenker97}.
Despite its widespread interest, a microscopic and quantitative modeling of
photoelasticity in amorphous silica is still lacking.

In this paper, we have computed the photoelastic tensor of crystalline and amorphous silica
by first principles within density functional perturbation theory (DFPT) 
\cite{baroni01} aiming at identifying the microscopic properties which rule the photoelastic
response in silica systems. Models of amorphous silica containing up to 162 atoms have been
generated by quenching from the melt in combined classical and Car-Parrinello molecular dynamics simulations.
Photoelastic coefficients of quartz have been computed previously within DFPT
by Detraux and Gonze \cite{detraux01}, who found good agreement with  experimental
data.
From the change of effective charges of oxygen upon strain, the latter authors have suggested 
that an important contribution to photoelasticity in quartz comes from the 
dependence of oxygen  polarizability on the deformation of the SiOSi angles upon strain.

In this paper we have enlarged the scope of the previous work by Detraux and Gonze
\cite{detraux01}
by computing the photoelastic coefficients of a-SiO$_2$ and of several crystalline phases
(quartz, $\alpha$-cristobalite, $\beta$-cristobalite)
at normal conditions and at high pressure from first principles.
Comparison with available experimental data and with calculations including a scissor correction to the 
electronic band gap,  let us conclude that the simple local density approximation to DFPT is suitable to 
well reproduce the experimental photoelastic tensor, despite a large error (10 $\%$) in the dielectric constants.
The ab-initio photoelastic coefficients of the crystalline phases of silica provided us with
a database over which a phenomenological 
model of the dielectric response has been  fitted.  This model aided us in the interpretation
of the ab-initio results on the  photoelastic properties of amorphous silica.
As described below, we have confirmed on a quantitative basis that the dependence of the 
anisotropic oxygen polarizability 
on the Si\^OSi angle
is the key parameter for modeling the dielectric and photoelastic properties of silica.
The paper is organized as follows. In section II we describe our computational framework. In section III we report
on a preliminary test of our framework on the photoelastic tensor of Si and MgO as prototypes of covalent and ionic
materials. 
In section IV we report our results on the photoelastic coefficients and structural response to strains of
several crystalline phases of silica: quartz at ambient conditions and high pressure, $\alpha$-cristobalite, and
$\beta$-cristobalite. In section V, we first report the details of the protocols used to generate the models of
a-SiO$_2$. Analysis of their structural and elastic properties are given. Then the  calculated photoelastic
tensor for a-SiO$_2$  is discussed and compared with experimental data. In section VI we present a phenomenological
model of the dielectric properties of silica that we have fitted on the dielectric tensor of
$\alpha$-cristobalite at different densities. Its transferability was proven by comparison with
the ab-initio database on the dielectric and photoelastic tensors of the crystalline and amorphous systems presented
in the preceding sections. Finally, section VII is devoted to discussion and conclusions.

\section{Computational details}
We have used density functional theory in the local density approximation (LDA) \cite{pz81},
norm conserving pseudopotentials  
for oxygen \cite{troullier} and  silicon \cite{gonze91}  and
plane-waves expansion of the Kohn-Sham (KS) orbitals up to a kinetic cutoff of 70 Ry, if not
specified otherwise.
The structure of the crystalline phases has been determined by optimizing 
all the  structural parameters. Monkhorst-Pack (MP)
\cite{MP} meshes have been used in the integration of the Brillouin Zone (BZ).
The calculations on the crystalline phases have been performed by  using the code
PWSCF \cite{Pwscf}. 
Models of a-SiO$_2$ have been generated by quenching from the melt in classical
molecular dynamics simulations using the empirical potential by van Beest {\sl et al.}
\cite{vanbeest} which has been shown to properly describe the structural and 
dynamical properties of amorphous silica \cite{tse92}. 
We have slightly modified the interatomic potential of ref.  \cite{vanbeest} for short
Si-O and O-O separations by adding a term of the form 
$V_{ij}=A_{ij}/r^{12}-B_{ij}/r^8$ with
 $A=6~eV\cdot{\rm\AA}^{12}$ and $B=20~eV\cdot\rm\AA^8$ for Si-O interactions
and   $A=24~eV\cdot\rm\AA^{12}$  and $B=180~eV\cdot\rm\AA^8$ for O-O interactions.
The modified potential turned out to be necessary to describe the 
high-temperature liquid (above 5000 K) where
ions with opposite charges may approach each other very closely and fall in the 
well of the Coulomb potential in the absence of the additional repulsion here 
introduced.
This new term does not change the structural properties of crystalline and amorphous
phases.
The simulations have been performed at constant volume.
Ab-initio annealing of the models at 600 $K$ for 0.75 ps 
has been then performed by Car-Parrinello \cite{cpmd}
molecular dynamics simulation as implemented in the code CPMD \cite{CPMD35}. 
 We have generated two amorphous models containing 81 and 162 atoms.
Different quenching and annealing protocols have been used as discussed in section V. 
Integration of the BZ has been restricted to the 
$\Gamma$ point only in the Car-Parrinello simulations. 
A fictitious electronic mass of 800 and a time step of 
 0.15 fs have been used.
In the dynamical Car-Parrinello simulations we have
used an ultrasoft pseudopotential \cite{vanderbilt} for oxygen and a lower cutoff of 27 Ry.
The structures generated by Car-Parrinello simulations have been then optimized with
norm-conserving  pseudopotentials to study the dielectric properties. 
We have computed the dielectric  and photoelastic tensors within density functional 
perturbation theory  \cite{baroni01}.
The photoelastic tensor $p_{ijkl}$ is defined by 
\begin{equation}
 \Delta\varepsilon^{-1}_{ij}=p_{ijkl}\eta_{kl}
 \end{equation}
where $\varepsilon_{ij}$ is the optical 
dielectric tensor and $\eta_{kl}$ is the strain tensor. 
Only the electronic contribution is included in $\varepsilon_{ij}$, also indicated as 
$\varepsilon^{\infty}$. Experimentally, it corresponds to the dielectric response 
measured for frequencies of the applied field much higher than
lattice vibrational frequencies, but lower than the 
frequencies of the electronic transitions. 
The photoelastic coefficients have been calculated by finite differences
from the  dielectric tensor
of  systems with strains from -2 $\%$
to +2 $\%$. The independent components of
$p_{ijkl}$ (21 in general) can be reduced by the symmetry of the system.
In quartz for instance there are only eight independent coefficients which can
be determined by applying three types of strain $\eta_{11}$, $\eta_{33}$ and $\eta_{23}$.
The components of the photoelastic
tensor will be expressed hereafter in  the compressed Voigt notation.

The dielectric tensor for the different strained configurations has been calculated
within DFPT \cite{baroni01}, as implemented in
the code PWSCF and PHONONS \cite{Pwscf}. 
However, the largest amorphous model considered here
which contains 162 atoms turned out intractable within the latter framework.
Therefore, we have also computed the dielectric tensor 
within the Berry phase (BP) approach to the  electronic polarization 
\cite{king,resta} as developed by Putrino {\sl et al.} \cite{putrino00}. 
In these latter calculations  we have restricted the BZ integration to the
$\Gamma$-point only. 
This latter technique 
 converges to the
correct DFPT result in the limit of large simulation cells or equivalent 
k-points mesh, although with a different pace  with respect to standard DFPT.
How large should the system be to reproduce the correct result within the BP approach
 has been investigated for $\beta$-cristobalite in section IVc below.
It turns out that for system size of the order of 192 atoms and the $\Gamma$-point only, 
the error in the
photoelastic coefficients of $\beta$-cristobalite are of the same order of the errors 
introduced by the LDA. 
Photoelasticity of the largest model of a-SiO$_2$
in section V has been addressed within the BP approach  as implemented in the code
CPMD \cite{CPMD35}.

The  exchange-correlation functionals available in literature (LDA and generalized gradient 
approximation (GGA)) usually underestimate the 
electronic band gap and  overestimate the electronic dielectric tensor up to
10-15 $\%$ \cite{detraux01}. 
This discrepancy can be corrected semi-empirically by applying a self-energy 
correction, also referred to as a scissor correction, which consists of  a rigid shift of
the conduction bands with respect to the valence bands \cite{scissor}.
This procedure has been used successfully to reproduce the photoelasticity of
Si \cite{levine92}, GaAs \cite{raynolds95} and quartz \cite{detraux01}.
However, as shown in section III and IV, it turns out that even within simple LDA, the error in the photoelastic
coefficients is  smaller than what expected on the basis of 
the error in the dielectric constant itself.
The scissor correction has thus been neglected in the calculations on amorphous silica and
on crystalline phases at high pressure. 
The calculations with the scissor corrections have been performed with the code ABINIT  \cite{abinit}.

\section{Test cases: silicon and MgO}
As test cases of our theoretical framework, prior  to the application to silica,
we have  computed the photoelastic coefficients of crystalline silicon and
MgO as prototypes of covalent and ionic systems, respectively.

\subsection{Silicon}

We have computed the photoelastic tensor for the conventional cubic supercell containing
eight atoms. 
The dielectric tensor has only one component and 
there are three photoelastic coefficients, $p_{11}$,
$p_{12}$ and $p_{44}$ independent by symmetry.
A  8x8x8 MP mesh over the BZ corresponding to
20 k-points in the irreducible wedge is used. The KS states are expanded in plane waves up to a kinetic
cutoff of 20 Ry.
The photoelastic coefficients, computed
at the theoretical equilibrium lattice constant ($a=5.377$ \AA), are
reported in table \ref{silicon} and compared with previous results obtained by Levine et al. \cite{levine92}.
These authors computed the photoelastic coefficients at the experimental lattice
parameter ($a=5.429$ \AA) within simple LDA and by adding a scissor operator,
corresponding to a rigid shift of 0.7 eV of the conduction band with respect to the valence band. 
Although the scissor correction is
crucial to reproduce the dielectric constant of silicon, it produces small changes in the
photoelastic tensor both  at the experimental and at the theoretical
equilibrium densities.
On the other hand, 
the  photoelastic coefficients are strongly dependent  
on the lattice constant (cfr.  table \ref{silicon}).
Better agreement with experiments 
is achieved when the theoretical lattice parameter is used.  
 This conclusion is also true for the calculation of the piezoelectric tensor in
other materials
\cite{degiro89,dalcorso93}.
\vspace{5 truecm}
\begin{table}[ht]
\begin{center}
\begin{tabular}{lccccc}
\hline\hline
                   & LDA              & LDA+sciss.        & LDA                 &  LDA+sciss.         & Exp. \\
                   & ($a_{theo}$)     &   ($a_{theo}$)    & ($a_{exp}$)         & ($a_{exp}$)         &      \\
                   & this work        & this work         &ref.\protect\cite{levine92} & ref.\protect\cite{levine92} & ref. \cite{expSi}\\
\hline
$\varepsilon$      & 13.20            & 11.38             &   -              &   11.4              &  11.4   \\
$p_{11}-p_{12}$    & -0.105           & -0.110            &-0.105               & -0.115              & -0.111$\pm$0.005 \\
$p_{11}+2p_{12}$   & -0.058           & -0.051            &-0.085               & -0.067              & -0.055$\pm$0.006 \\
$p_{11}$           & -0.090           & -0.090            &-0.098               & -0.099              &  -0.094$\pm$0.005 \\
$p_{12}$           &  0.016           &  0.020            & 0.007               &  0.016              &  0.017$\pm$0.001 \\
$p_{44}$           & -0.052           &   -               &-0.045               & -0.049              & -0.051$\pm$0.002 \\
\hline\hline
\end{tabular}
\end{center}
\caption{Dielectric constant and photoelastic coefficients of crystalline silicon.
Results obtained by simple LDA and by 
adding a scissor correction of 0.7 eV are compared. Previous theoretical results by
Levine {\sl et al.} \protect\cite{levine92} are also reported.} 
\label{silicon}
\end{table}

\subsection{MgO}

MgO crystallizes in a NaCl-type  lattice. 
We have computed the photoelastic coefficients for a conventional cubic
supercell containing eight atoms.
A norm conserving pseudopotential of Car-Von Barth type \cite{vbc} including non-linear-core corrections 
\cite{nlcc} for Mg and
a 8x8x8 MP mesh for the BZ integration have been used. The equilibrium lattice parameter
has been obtained by fitting the equation of state with a Murnaghan function \cite{murna} which
yields $a= 4.197$ \AA\ ($a_{exp}=4.2017$ \AA) and a bulk modulus $B= 185$ GPa
($B_{exp}=160$ GPa). 
The dielectric constant is, as usual, overestimated ($\sim 5 \%$), but 
the calculated photoelastic coefficients reported in table
\ref{mgo} are in good agreement with experimental data \cite{cardona59,vedam66}.
The spread in the  experimental data is quite large, as they have not been obtained by 
the Brillouin scattering method but by a hydrostatic 
pressure method which is subject to larger uncertainties. 
\begin{table}[ht]
\begin{center}
\begin{tabular}{lccc}
\hline\hline
                   & LDA ($a_{theo}$) & Exp. \cite{cardona59} &  Exp. \cite{vedam66} \\
\hline
                   $\varepsilon$      &  3.16    &  2.92  &  3.02    \\
                   $p_{11}-p_{12}$    & -0.26    & -0.24  & -0.248    \\
                   $p_{11}$           & -0.31    & -0.30  & -0.259    \\
                   $p_{12}$           & -0.05    & -0.08  & -0.011     \\
                   $p_{44}$           & -0.075   &  -  & -0.096    \\
\hline\hline
\end{tabular}
\end{center}
\caption{Dielectric constant and photoelastic coefficients of MgO.}
\label{mgo}
\end{table}

\section{Crystalline silica}
In the next sections we report the calculated photoelastic coefficients of three crystalline
phases of silica at normal conditions and at high pressure: $\alpha$-quartz, $\alpha$-cristobalite, and
$\beta$-cristobalite. The collected set of data have then been used to fit the
phenomenological model of photoelasticity  discussed in section VI below.

\subsection{$\alpha$-Quartz}

The photoelastic coefficients of quartz have been recently computed within DFPT by
Detraux and Gonze \cite{detraux01}. This phase has thus been  used as a test case for our
theoretical framework on silica systems. The unit cell of quartz is hexagonal (space group $P3_121$)
and contains nine atoms of which  two  are independent by symmetry
in the positions Si $(u,0,0)$ and O $(x,y,z)$. We have used a 3x3x2 MP mesh in the BZ integration although
a single special point is enough to achieve convergence in the structural properties \cite{detraux01}.
The optimized lattice parameters ($a$, $c/a$) and internal 
structural parameters are reported in table
\ref{quartz}. Actually, the agreement with experiments on structural parameters is
fair, but somehow worse than usual for  the Si$\widehat{\rm O}$Si angle and 
the lattice parameter.
These discrepancies have to be mainly ascribed to the choice of the LDA for the
exchange and correlation functional which commonly causes 
 an underestimation of the equilibrium volume. 
It was demonstrated that GGA is necessary to reproduce the correct
energy differences among the phases of silica \cite{hamann96}, however the equilibrium lattice parameters and
the Si-$\widehat{\rm O}$-Si angle are only slightly improved for $\alpha$-quartz 
(agreement with  experimental data is still within 2$\%$) and 
Si-O distances slightly worsen (see table 1 of ref.\cite{hamann96} for comparison). 
The small differences between our results and those reported in ref.\cite{detraux01} depend
on the different choice of the pseudopotentials. 
\begin{table}[!ht]
\begin{center}
\begin{tabular}{lcccc}
\hline\hline
                   &  This work  & LDA                  &  LDA + scissor & experiment \\
                   &             & Ref. \cite{detraux01}&  Ref. \cite{detraux01} &  Ref. \cite{quartzexp,narasim69}\\
\hline
 a             & 4.805 & 4.815 &   & 4.913 \\
c/a                & 1.101 & 1.105 &   & 1.100 \\
 u                 & 0.460 & 0.461 &   & 0.465 \\
 x                 & 0.408 & 0.410 &   & 0.415 \\
 y                 & 0.283 & 0.281 &   & 0.272 \\
 z                 & 0.105 & 0.108 &   & 0.120 \\
Si-O (1)      & 1.605 & 1.602 &   & 1.605 \\
Si-O (2)      & 1.610 & 1.611 &   & 1.614 \\
Si-$\widehat{\rm O}$-Si & 137.7 & 139.1 &   & 143.7 \\
O-$\widehat{\rm Si}$-O  & 108.33 & 108.07 & &  108.37   \\
                        & 111.39 & 110.97 & &  110.88   \\
                        & 108.39 & 109.36 & &  108.37   \\
                        & 109.11 & 109.43 & &  110.41   \\
\hline
$\varepsilon_{11}$ & 2.590 & 2.538 &  2.353  & 2.356 \\
$\varepsilon_{33}$ & 2.620 & 2.569 &  2.385  & 2.383 \\
\hline
$p_{11}$           & 0.17  & 0.16  &  0.17  & 0.16  \\
$p_{12}$           & 0.23  & 0.23  &  0.23  & 0.27  \\
$p_{13}$           & 0.24  & 0.24  &  0.25  & 0.27  \\
$p_{33}$           & 0.085 & 0.080 &  0.11  & 0.10  \\
$p_{31}$           & 0.25  & 0.25  &  0.27  & 0.29  \\
$p_{14}$           &-0.031 & -0.023& -0.03  & -0.03  \\
$p_{41}$           &-0.040 & -0.034& -0.028 & -0.047 \\
$p_{44}$           &-0.056 & -0.062& -0.061 & -0.079 \\
\hline\hline
\end{tabular}
\end{center}
\caption{Structural parameters, dielectric tensor and photoelastic coefficients of $\alpha$-quartz as
computed in this work and measured experimentally \protect\cite{quartzexp,quartzexp2,narasim69}.
Previous results by Detraux and Gonze \protect\cite{detraux01} with the scissor correction
are also reported for sake of comparison. Angles are in degree and lengths in \AA.}
\label{quartz}
\end{table}

The calculated dielectric constants and independent photoelastic coefficients
are collected in table \ref{quartz}. 
Only three independent strains ($\eta_{11}$, $\eta_{33}$, and
$\eta_{23}$) are necessary to compute all the independent $p_{ij}$.
The theoretical $p_{ij}$ are in good agreement with previous results obtained by Detraux and Gonze \cite{detraux01}
by adding the scissor correction and with experiments.  Although the scissor operator is
necessary to reproduce the dielectric constant,  its effect
on the photoelastic coefficients is less important. 
The scissor operator has been adjusted to 2.1 eV in ref. \cite{detraux01}
to better reproduce the dielectric constants of quartz.  

 To identify the main structural changes upon strain, we have computed
the response of the structural parameters (bondlengths and angles) to strain of $\pm 2\%$.
Results are summarized in table \ref{struct}. The stiffness of the Si-O bonds 
causes the structure to respond to strain mainly by cahnges of the bond angles. 
The Si-$\widehat{\rm O}$-Si angles undergo larger changes than the tetrahedral O-$\widehat{\rm Si}$-O angles
due to the strongly directional character of the sp$^3$ orbitals of  silicon.
\begin{table}[!ht]
\begin{center}
\begin{tabular}{lccc}
\hline\hline
           &  $\partial/\partial \eta_{11}$ & $\partial/\partial \eta_{33}$ \\
\hline
Si-O (1) (\AA)         &  0.036    &  0.102            \\ 
Si-O (2) (\AA)         &  0.024    &  0.090             \\
Si-$\widehat{\rm O}$-Si (1)&  130.6    &  77.7              \\  
Si-$\widehat{\rm O}$-Si (2)&   30.5     &   -                \\
Si-$\widehat{\rm O}$-Si (3)&  160.4    &   -               \\
O-$\widehat{\rm Si}$-O (1)   &   77.2   &  47.5        \\
O-$\widehat{\rm Si}$-O (2)   &  -33.9    & -0.6        \\
O-$\widehat{\rm Si}$-O (2')  &  -19.0    &   -         \\
O-$\widehat{\rm Si}$-O (3)   &   36.0    & -51.7        \\
O-$\widehat{\rm Si}$-O (3')  &    -12.9  &   -         \\
O-$\widehat{\rm Si}$-O (4)   & -15.7     &  57.4        \\
\hline\hline
\end{tabular}
\end{center}
\caption{The response of Si-O bondlength and Si-$\widehat{\rm O}$-Si angles of $\alpha$-quartz
         to $\eta_{11}$,  $\eta_{33}$ and $\eta_{23}$ strains. Symmetry breaking due to 
          $\eta_{11}$ strains  generates three independent  Si-$\widehat{\rm O}$-Si angles.}
\label{struct}
\end{table}

The dielectric tensor and photoelastic coefficients have been  calculated at high
pressure as well. Table \ref{pression} reports the structural and dielectric properties of $\alpha$-quartz at
3 and 7 GPa, well below the transition pressure to the monoclinic phase (21 GPa)
\cite{hemley}.
The variation of the equilibrium volume and of the internal structural parameters of 
$\alpha$-quartz as a function of pressure is in  good agreement with experiments 
\cite{quartzP}. The Si-$\widehat{\rm O}$-Si angle undergoes larger changes upon compression than
the O-$\widehat{\rm Si}$-O angles, while the Si-O distances are nearly unaffected.

\begin{table}[!ht]
\begin{center}
\begin{tabular}{lccc}
\hline\hline
 & 0 GPa   &  3 GPa  &  7 GPa  \\
\hline
 a                 & 4.805 &  4.661 & 4.517 \\
c/a                & 1.101 &  1.120 & 1.142 \\
V/V$_0$            &  1    & 0.93 (0.93) & 0.86 (0.87) \\
 u                 & 0.460 &  0.450 & 0.441 \\
 x                 & 0.408 &  0.400 & 0.391 \\
 y                 & 0.283 &  0.299 & 0.311 \\
 z                 & 0.105 &  0.095 & 0.088 \\
Si-O     & 1.605 (1.605) &  1.600 (1.602) & 1.593 (1.599) \\
Si-O     & 1.610 (1.614) &  1.611 (1.613) & 1.611 (1.614) \\
Si-$\widehat{\rm O}$-Si & 137.7 (143.7) &  131.8 (139.9) & 126.8 (133.3) \\
O-$\widehat{\rm Si}$-O  & 108.3     & 106.2  & 103.7 \\
O-$\widehat{\rm Si}$-O  & 111.4     & 113.6  & 115.9 \\
O-$\widehat{\rm Si}$-O  & 108.3     & 107.5  & 106.6 \\
O-$\widehat{\rm Si}$-O  & 109.1     & 108.6  & 108.3 \\
\hline
$\varepsilon_{11}$ & 2.590 &  2.694 & 2.812 \\
$\varepsilon_{33}$ & 2.620 &  2.737 & 2.865 \\
\hline
$p_{11}$           & 0.170 &  0.174 & 0.219 \\
$p_{12}$           & 0.227 &  0.214 & 0.209 \\
$p_{13}$           & 0.245 &  0.215 & 0.199 \\
$p_{33}$           & 0.085 &  0.067 & 0.051 \\
$p_{31}$           & 0.253 &  0.251 & 0.254 \\
$p_{44}$           &-0.056 & -0.055 &-0.053 \\
\hline\hline
\end{tabular}
\end{center}
\caption{Structural parameters, dielectric and photoelastic tensor of $\alpha$-quartz at high
pressure. Experimental data from Ref.\cite{quartzP} are given in parenthesis. Angles are in degree and lengths in \AA.}
\label{pression}
\end{table}
We are not aware of any experimental
data on the photoelastic coefficients of quartz at high pressure.

The dependence on pressure of the response of structural parameters to strain is reported in table \ref{strp}.
By inspecton of table \ref{strp} we can conclude
that at high pressure the Si-O bonds become even stiffer.
However, since the flexibility of the Si-$\widehat{\rm O}$-Si angle is reduced,
also the deformation of the tetrahedral unit becomes sizable.
\begin{table}[!ht]
\begin{center}
\begin{tabular}{lccc}
\hline\hline
$\partial/\partial\eta_{33}$ &  0 GPa    &  3 GPa   &  7 GPa   \\
\hline
Si-O (1)          & 0.102     & 0.078    &  0.031   \\
Si-O (2)          & 0.090     & 0.013    & -0.005   \\
Si-$\widehat{\rm O}$-Si    & 77.7      & 69.4     &  61.7    \\
O-$\widehat{\rm Si}$-O (1)  &  47.5     & 0.81     & -19.6    \\
O-$\widehat{\rm Si}$-O (2)  &  -0.70    & 20.3     &  17.6    \\
O-$\widehat{\rm Si}$-O (3)  & -51.7     & -41.8    & -34.9    \\ 
O-$\widehat{\rm Si}$-O (4)  &  57.4     & 42.4     &  54.6    \\
\hline\hline
\end{tabular}
\end{center}
\caption{The response of the structural parameters of $\alpha$-quartz to 
         the $\eta_{33}$ strain at different pressures. Angles are in degree and lengths in \AA.}
\label{strp}
\end{table}
The data reported so far are not sufficient to establish a correlation between structural and photoelastic
properties, however these results support the idea that bond angles (mainly Si-$\widehat{\rm O}$-Si)
play a key role in determining the 
photoelastic coefficients of quartz.

In order to identify the structural response of silica polymorph to strain, it is useful to consider 
this system as a network of SiOSi units. 
The structural response to strain can then be expressed in terms of the length and orientation of
the vector distance  Si-Si between the two silicon atoms of the SiOSi unit
which depends on the
Si-$\widehat{\rm O}$-Si bond angle. The orientation of the SiOSi units can be valuated by computing the 
projection of the Si-Si distances on the Cartesian axis.  
This analysis (tab. \ref{psisiq})
shows that a tensile strain increases the alignment of the Si-Si vectors along the strain axis.
\begin{table}[!ht]
\begin{center}
\begin{tabular}{ccccc}
\hline\hline
                  &   &  0 GPa    &  3 GPa   &  7 GPa   \\
\hline
         & P$_x$(Si-Si)         & 1.690     & 1.554    &  1.505   \\
$\partial/\partial\eta_{11}$ & P$_y$(Si-Si)         & 0.000     &-0.024    & -0.006    \\
         & P$_z$(Si-Si)         & 0.000     &-0.008    & -0.012    \\
\hline
         & P$_x$(Si-Si)         & -0.450    &-0.392    & -0.416    \\
$\partial/\partial\eta_{33}$ & P$_y$(Si-Si)         &  0.000    & 0.000    & -0.250   \\
         & P$_z$(Si-Si)         & 1.762     & 1.595    &  1.718   \\
\hline\hline
\end{tabular}
\end{center}
\caption{The derivatives of the projection of the Si-Si vectors along the direction $i$ (P$_i$(Si-Si))
         with respect to strains $\eta_{11}$ and $\eta_{33}$ in $\alpha$-quartz at different pressures.}
\label{psisiq}
\end{table}

\subsection{$\alpha$-Cristobalite}

The $\alpha$-cristobalite crystal has tetragonal symmetry (space group $P4_12_12$).
The unit cell contains 12 atoms of which two are independent by symmetry at the positions
Si $(u,u,0)$ and O $(x,y,z)$. 
The optimized structural parameters
of $\alpha$-cristobalite are compared with
experiments in table \ref{cristo1}.
A 3x3x2 MP mesh has been used in the BZ integration.
The experimental value of Si-$\widehat{\rm O}$-Si angle (144.7$^o$) which is suggested to be the structural
parameter which mostly
influences the photoelastic response \cite{detraux01} is very similar to that of quartz
(143.9$^o$).
Conversely, $\alpha$-cristobalite  and quartz differ in
density (2.36 g/cm$^3$ and 2.65 g/cm$^3$ for $\alpha$-cristobalite and $\alpha$-quartz, respectively)
and in the ring topology (6- and 8-membered rings in quartz and only 6-membered rings
in $\alpha$-cristobalite). 
\begin{table}[ht]
\begin{center}
\begin{tabular}{lcc}
\hline\hline
                   & This work   &  Exp. (Ref.\cite{pluth85})  \\
\hline
 a                 & 4.856 &         4.957 \\
c/a                & 1.381 &         1.390 \\
 u                 & 0.313 &         0.305 \\
 x                 & 0.234 &         0.238 \\
 y                 & 0.129 &         0.111 \\
 z                 & 0.191 &         0.183 \\
Si-O (1)      & 1.603 &         1.602 \\
Si-O (2)      & 1.605 &         1.617 \\
Si-$\widehat{\rm O}$-Si   & 139.8         & 144.7 \\
O-$\widehat{\rm Si}$-O (1)   &  107.53  &  108.08 \\
O-$\widehat{\rm Si}$-O (2)   &  109.36  &  109.57 \\
O-$\widehat{\rm Si}$-O (3)   &  110.35  &  109.92 \\
O-$\widehat{\rm Si}$-O (4)   &  111.71  &  111.27 \\
\hline\hline
\end{tabular}
\end{center}
\caption{Calculated and experimental structural parameters of $\alpha$-cristobalite. 
         Angles are in degree and length in \AA.}
\label{cristo1}
\end{table}

The dielectric tensor has two independent components
($\varepsilon_{11}=\varepsilon_{22}$ and $\varepsilon_{33}$).  The independent
photoelastic coefficients are seven and can be obtained by applying of four strains:
$\eta_{11}$, $\eta_{33}$, $\eta_{12}$ and $\eta_{23}$.
The theoretical dielectric and photoelastic tensors are reported in table
\ref{cristo2}. The results obtained by including a scissor correction 
 equal to that used for quartz (2.1 eV) \cite{detraux01} are also reported. 
The scissor correction largely improves the
dielectric constants, but produces smaller changes in the photoelastic coefficients as
occurs in quartz.
Hereafter, the scissor correction will thus be neglected in the calculation of the
photoelastic tensor for other crystalline and amorphous phases of SiO$_2$.

We are not aware of any experimental measurement of the photoelastic coefficients
in $\alpha$-cristobalite.
\begin{table}
\begin{center}
\begin{tabular}{lcc}
\hline\hline
         & LDA & LDA+scissor \\
\hline
$\varepsilon_{11}$ & 2.363  & 2.206 (2.211) \\
$\varepsilon_{33}$ & 2.358  & 2.196 (2.202) \\
\hline
$p_{11}$           & 0.218  & 0.225  \\
$p_{12}$           & 0.244  & 0.248    \\
$p_{13}$           & 0.293  & 0.299   \\
$p_{33}$           & 0.152  & 0.162  \\
$p_{31}$           & 0.294  & 0.298  \\
$p_{44}$           &-0.094  &-0.090 \\
$p_{66}$           &-0.068  &-0.066   \\
\hline\hline
\end{tabular}
\end{center}
\caption{Dielectric and photoelastic tensors of $\alpha$-cristobalite within
the simple LDA and LDA supplemented by the scissor correction (2.1 eV). 
Experimental values for $\varepsilon_{11}$ and
$\varepsilon_{33}$ are reported in parenthesis \cite{cristoexp}.}
\label{cristo2}
\end{table}

The differences in  density and  topology of quartz and $\alpha$-cristobalite
determine a different response of the 
structural parameters to strain (results are displayed in tab. \ref{a_str}). 
\begin{table}[!ht]
\begin{center}
\begin{tabular}{lccc}
\hline\hline
                       &  $\partial/\partial \eta_{11}$ & $\partial/\partial \eta_{33}$ & $\partial/\partial \eta_{23}$ \\
\hline
Si-O (1)          &  0.030    &   -0.011 & $\pm$0.006 \\
Si-O (1')          & -0.019    &     -    & $\pm$0.125 \\
Si-O (2)          & -0.079    &   -0.008 & $\pm$0.015 \\
Si-O (2')          &  0.031    &     -    & $\pm$0.172 \\
Si-$\widehat{\rm O}$-Si (1) & -31.2     &   70.6   & $\pm$33.3 \\
Si-$\widehat{\rm O}$-Si (2) & 186.7     &    -     & $\pm$37.2 \\
O-$\widehat{\rm Si}$-O (1)   &  9.98     &   -35.3  & $\pm$30.5 \\
O-$\widehat{\rm Si}$-O (1')  & 68.2      &    -     & $\pm$18.2 \\
O-$\widehat{\rm Si}$-O (2)   & -14.9     &  3.81    & $\pm$41.9 \\
O-$\widehat{\rm Si}$-O (3)   & -16.9     &  14.5    & $\pm$20.5 \\
O-$\widehat{\rm Si}$-O (3')  & -18.4     &   -      & $\pm$54.4 \\
O-$\widehat{\rm Si}$-O (4)   & -29.1     &  38.3    & $\pm$24.8 \\
\hline\hline
\end{tabular}
\end{center}
\caption{The response of Si-O bondlength and bond angles of $\alpha$-cristobalite
         to $\eta_{11}$,  $\eta_{33}$ and $\eta_{23}$ strains. 
          The number of different Si-O bondlengths, 
          Si-$\widehat{\rm O}$-Si and O-$\widehat{\rm Si}$-O angles increase upon application of the
	   $\eta_{11}$ strain and further double under application of thhe $\eta_{23}$ strain.
	   Angles are in degree and lengths in \AA.}
\label{a_str}
\end{table}
In $\alpha$-cristobalite bond lengths are affected by strain even less than in quartz
and the  distortion of the tetrahedral bond angles is smaller. 
In fact, $\alpha$-cristobalite has a more open structure than quartz and it is able to accommodate 
strains mainly by a rotation of the tetrahedra.

The derivatives 
of the projection of the Si-Si vectors along the Cartesian axes with respect to strain 
are reported in Tab.\ref{asisiq}.
The alignment of the Si-O-Si units along the strain axis 
observed in $\alpha$-quartz occurs also in $\alpha$-cristobalite for strain along the 
main symmetry axis ($\eta_{33}$), but it
is much smaller for the $\eta_{11}$ strain.
\begin{table}[!ht]
\begin{center}
\begin{tabular}{ccc}
\hline\hline
                     &  $\partial /\partial\eta_{11}$ & $\partial /\partial\eta_{33}$ \\
\hline
  P$_x$(Si-Si)         & 0.190     &-0.675  \\
  P$_y$(Si-Si)         &-0.226     &-0.675  \\
  P$_z$(Si-Si)         & 0         & 1.676  \\
\hline\hline
\end{tabular}
\end{center}
\caption{The derivatives of the projection of the Si-Si vectors along the direction $i$ (P$_i$(Si-Si))
         with respect to strains $\eta_{11}$ and $\eta_{33}$ in $\alpha$-cristobalite.}
\label{asisiq}
\end{table}

\subsection{$\beta$-Cristobalite}

The $\beta$-cristobalite phase can be obtained by heating $\alpha$-cristobalite
above 270 $^o$C at normal pressure \cite{wright}.
The space group $Fd3m$ assigned experimentally to $\beta$-cristobalite
has been interpreted  in the past as an average structure composed of small domains with six possible orientations
of the structure with $I\bar{4}2d$ symmetry \cite{wright}.
More recent NMR data have shown that the coexistence of different domains of lower symmetry is in
fact dynamical
and not static as due to domains with fixed orientation \cite{betadyn}.
However, the $\beta$-cristobalite structure with $I\bar{4}2d$ symmetry 
is locally stable
at low temperature (0-300 K) in the small simulation cells with periodic boundary conditions used here.
Its properties
can then be studied theoretically, also to get insight into the properties of
amorphous silica. In fact, $\beta$-cristobalite is the crystalline phase of silica with density
(2.18 g/cm$^3$) and refractive index closest to those of amorphous silica.
We have  modeled
$\beta$-cristobalite in a cubic supercell containing 8 formula units with  $I\bar{4}2d$ symmetry.
We optimized the internal coordinates at the experimental density of 2.18 g/cm$^3$ ($a=7.13$ \AA).
A 4x4x4 MP mesh has been used in the BZ integration.
The internal structural parameters of $\beta$-cristobalite are assigned by the Si\^OSi angle, two
O\^SiO angles and the Si-O bond length. The calculated values are 142.9$^o$ (Si\^OSi), 107.5$^o$ and
113.6$^o$ (O\^SiO) and 1.606 \AA\ (Si-O) in good agreement with the experimental values 146.7$^o$,
107.8$^o$ and 112.8$^o$,  and 1.611 \AA\, respectively \cite{wright}.

The dielectric tensor has two independent components
$\varepsilon_{11}=\varepsilon_{22}$ and $\varepsilon_{33}$. The calculated dielectric constants and
photoelastic coefficients are reported in table \ref{betacris}. 

In the perspective to study
the dielectric response of large models of amorphous silica, we have checked the convergence of
the photoelastic coefficients as a function of the size of the mesh in the BZ integration
for $\beta$-cristobalite.
Table \ref{betacris} reports the results obtained by using the $\Gamma$-point only of the
24-atoms supercell and a mesh of 8 k-points in the first BZ which corresponds to the $\Gamma$-point only
of a cubic supercell containing 192 atoms. 
The results with a mesh of 32 k-points are the ``exact'' values at convergence in the BZ integration.
The dielectric and photoelastic tensor are
already converged by using the $\Gamma$-point of  the 192-atoms supercell.
We have also computed the dielectric tensor and photoelastic coefficients of the 192-atoms
supercell of $\beta$-cristobalite within the
Berry phase approach to electronic polarization,
 as developed by Putrino {\sl et al.} \cite{putrino00}. This latter calculation
converges to the
correct DFPT result, although with a different pace with the cell size by using the
 $\Gamma$-point only in the BZ integration, or, equivalently, with the $\vec{k}$-point mesh \cite{umari02}.
In fact, calculations within DFPT and BP with the same $\vec{k}$-points mesh give different results
and the discrepancy decreases by increasing the $\vec{k}$-points mesh (or, equivalently, the
cell dimension). In table \ref{betacris} the dielectric and photoelastic
tensor of $\beta$-cristobalite calculated within DFPT and BP are compared. 
 The BP approach has been applied to a $\beta$-cristobalite supercell containing 192 atoms, by restricting
the BZ integration to the $\Gamma$-point only, which is equivalent to the 8 $\vec{k}$-points mesh of the
24 atoms supercell. 
However, the DFPT results with 8 k-points (24-atoms cell) are different from the BP
results on an equivalent k-point mesh ($\Gamma$-point and 192-atoms cell).
The BP approach convereges 
 converges more slowly with cell size (or equivalently with the $\vec{k}$-points mesh) than 
DFPT. 
Moreover, the convereged value of $\tens{\varepsilon}$ is approached from below within 
the BP approach.
This effect is responsible for the difference in dielectric constant of the a-SiO$_2$ reported here (see infra)
 for a 81 supercell and a 162-atoms cell and for a 72-atoms 
cell recently considered in ref. \cite{umari02}.

From table \ref{betacris} it turns out that
for system size of the order of 192 atoms the errors in the
photoelastic coefficients due to the use of BP approach at 
$\Gamma$-point are of the same order of the errors
introduced by the LDA itself.
Since the BP approach is computationally less demanding that DFPT as implemented in the codes
CPMD and PWSCF, respectively, we have used the former approach for the largest amorphous model described in the
next section.
\begin{table}
\begin{center}
\begin{tabular}{lcccc}
\hline\hline
                   & $\Gamma$ & 8 {\bf k} & 32 {\bf k} points & BP 192 atoms \\\hline
$\varepsilon_{11}$ & 2.337  & 2.281  & 2.282 & 2.181 \\
$\varepsilon_{33}$ & 2.310  & 2.251  & 2.252 & 2.161 \\
\hline
$p_{21}$           & 0.297  & 0.291  & 0.292 & 0.266 \\
$p_{22}$           & 0.131  & 0.124  & 0.119 & 0.165 \\
$p_{23}$           & 0.242  & 0.269  & 0.267 & 0.238 \\
\hline\hline
\end{tabular}
\end{center}
\caption{Dielectric and photoelastic tensors of $\beta$-cristobalite calculated at convergence
with respect to BZ integration (MP 4x4x4 mesh corresponding to
32 $\vec{k}$-points in the irreducible wedge) and with two other coarser meshes ($\Gamma$-point only and 8
 $\vec{k}$-points in the IBZ, see text). The cubic unit cell containing 24 atoms has been used.
The results obtained with the code CPMD \cite{CPMD35} and the  Berry-phase approach of
 ref. \cite{putrino00} is reported for the 192-atoms supercell at the $\Gamma$-point only.}
\label{betacris}
\end{table}

\section{Amorphous silica}

We have generated several models of amorphous silica by quenching from the melt in
classical MD simulations
using the empirical potential by van Beest {\sl et al.}
\cite{vanbeest}
as described in section II. We have considered simulation cells
of two sizes: 81 and 162 atoms.
We first report on the details of the quenching protocols and on the
structural properties of the resulting amorphous models before 
discussing the dielectric properties of a-SiO$_2$.

\subsection{Structural properties}

The generation of amorphous models by quenching from the melt within Car-Parrinello simulations
poses some unavoidable restrictions on the cell size and the quenching time, 
which, in turn, negatively affect the 
quality of the  models produced. 
The main problem is that
a very high quenching rates 
(in ref. \cite{sarnthein95} quenching rate was 10$^{16}$ K/s), 
 prevents an adequate  relaxation  which
freezes the  concentration of small rings (three and four membered) to a value much higher than those
expected in the experimental samples \cite{sarnthein95}. A further limitation is given by the 
small size of the systems that can be treated by ab-initio simulations (in ref.
\cite{sarnthein95} the simulation supercell contains 72 atoms). 

\begin{figure}[!ht]
\centerline{\epsfxsize= 7. truecm
\epsffile{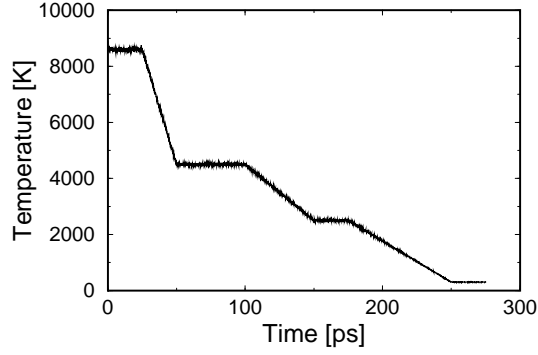}}
\caption{Quenching protocol adopted for the 162-atoms amorphous model.}
 \label{cicloT}
\end{figure}

In order to partially overcome these limitations, we have adopted a combined classical and ab--initio
simulation scheme, that has been already reported to produce models of glassy SiO$_2$
with satisfactory structural and electronic properties, when  compared to 
fully ab-initio models and to available experimental data \cite{benoit00}.
Within this approach, models of a-SiO$_2$ are generated by quenching 
from the melt by classical MD  with the BKS potential at low quenching rate. 
The amorphous models thus obtained are then annealed by Car--Parrinello MD.
Starting configurations of liquid silica have
been obtained from  unstable simple cubic crystal which rapidly reaches 8500 K.
For both the 81- and 162-atoms models we have used an initially cubic box with
 density of 2.21 g/cm$^3$.
After 25 ps at 8500 K,  the high temperature liquid is cooled to about 4000 K in 25 ps. 
The temperature at which atomic diffusion freezes over our simulation scale is much larger
($\sim$3500 K) than the experimental vitreous temperature ($\sim$1700 K). In order to obtain a 
good amorphous model, we have equilibrated the liquid at a temperature 
slightly above 3500 K and then quenched it slowly. 
The quenching protocol used to generate the 162-atoms model and a first 81-atoms model
is shown in Fig. \ref{cicloT}. 
It turns out that the quenching rate of Fig. \ref{cicloT} is still too fast to correctly reproduce
the structural properties of amorphous silica in the small system (81 atoms) which in fact
appears to be strongly anisotropic.
This problem is less severe for the larger system (162 atoms) which
can escape from high energy local minima visited along the quenching.
We have then generated another smaller model (81-atoms) with a much lower quenching rate by
equilibrating the system for 5 ns at 3800 K and quenching to 300 K in 2.5 ns 
(corresponding to a quenching rate of 1.4$\cdot$10$^{12}$ K/s).
The 81-atoms system referred to hereafter corresponds to this second quenching protocol.

\begin{figure}[!ht]
\centerline{\epsfxsize= 8.4 truecm
\epsffile{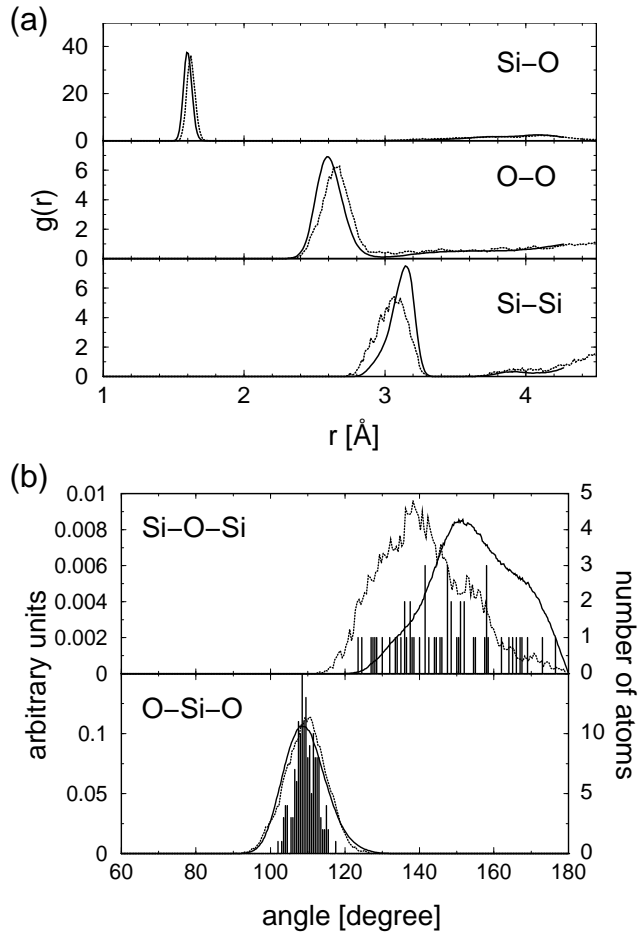}}
\caption{(a) Partial pair correlation functions  and (b) bond angle distributions  for
         the 81-atoms model of a-SiO$_2$, computed within classical MD (solid lines) and  CPMD (dotted lines) at 300 K.
	 Histograms in panel (b) refer to
         the   structure optimized ab-initio at the theoretical equilibrium density and at zero temperature.}
 \label{struct81}
\end{figure}

Ab-initio annealing for 1.0 ps at 600 $K$
has been then performed by CPMD. For the smaller model, 
we have computed structural properties over a dynamical ab-initio microcanonical run 
at 300 K, 0.6 ps long. The pair correlation functions and the 
bond angle distributions of the ab-initio and classical models at 300 K are compared 
in fig. \ref{struct81}. 

\begin{figure}[!ht]
\centerline{\epsfxsize= 8.4 truecm
\epsffile{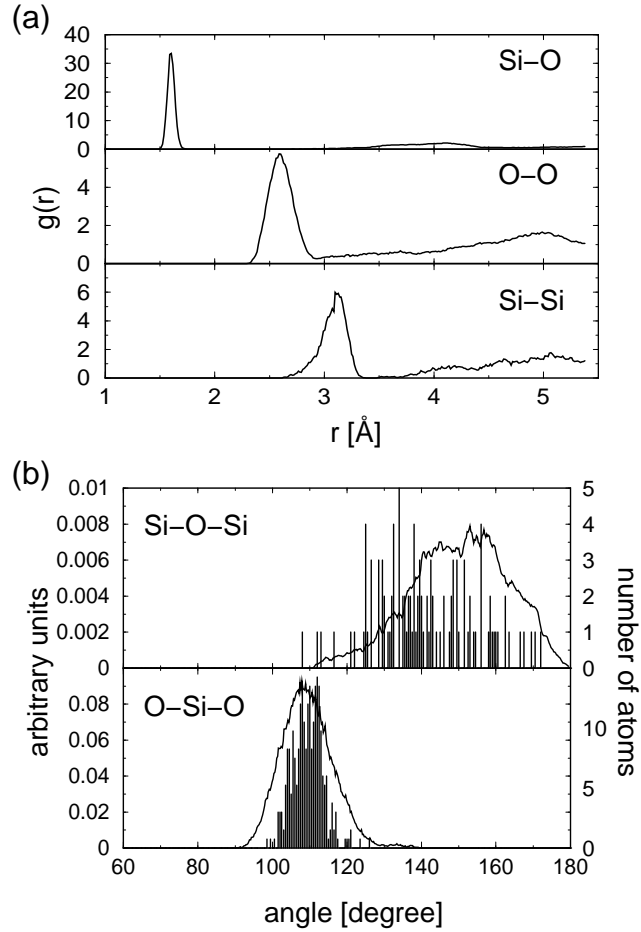}}
 \caption{(a) Partial pair correlation functions  and (b) bond angle distributions  for
	  the 162-atoms model of a-SiO$_2$, computed within classical MD  at 300 K.
 Histograms in panel (b) refer to
	  the   structure optimized ab-initio at the theoretical equilibrium density and at zero temperature.}
 \label{struct162}
\end{figure}

\begin{figure}[!ht]
  \begin{center}
  \centerline{\epsfxsize= 4.3 truecm\epsffile{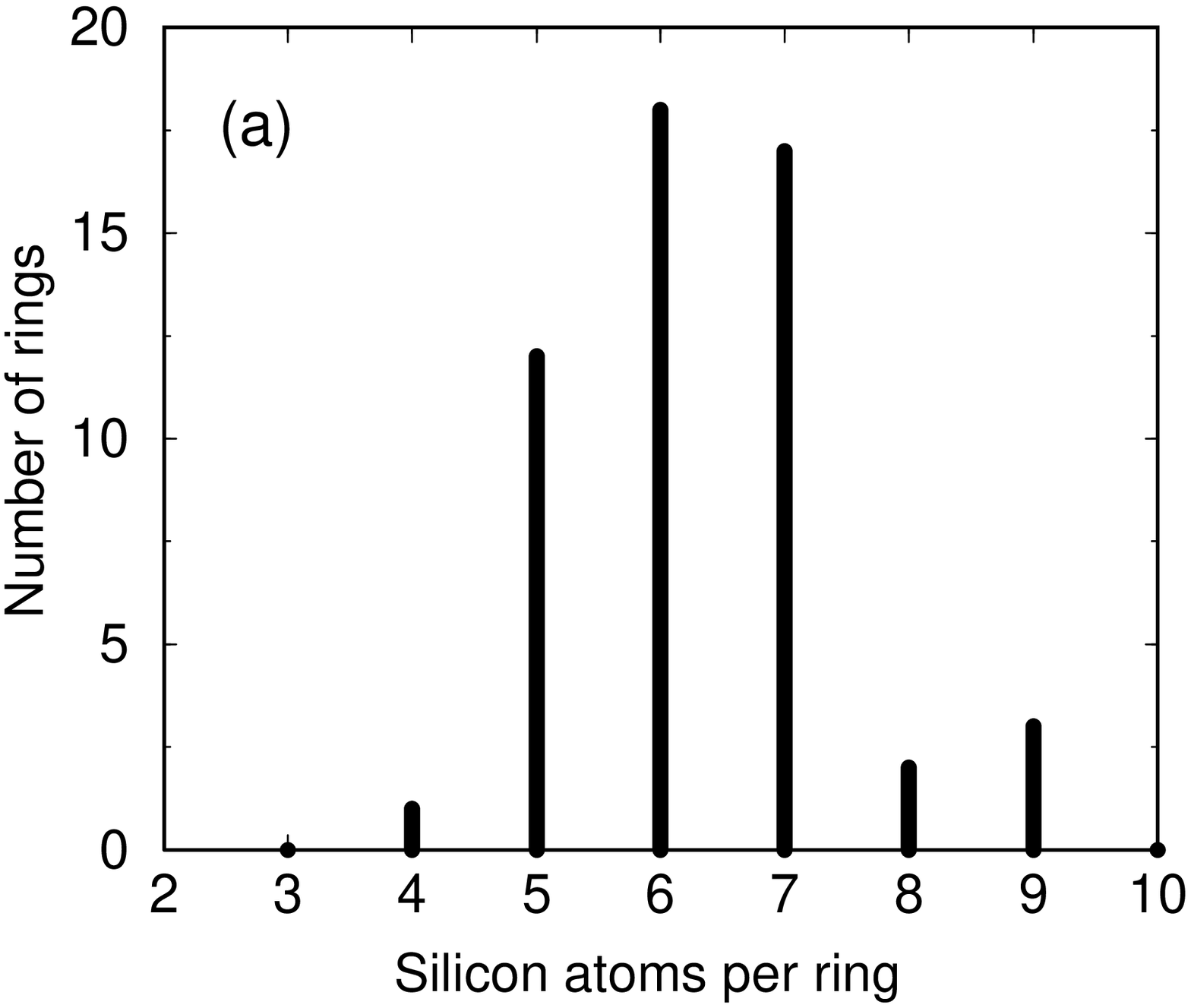}\epsfxsize= 4.3 truecm\epsffile{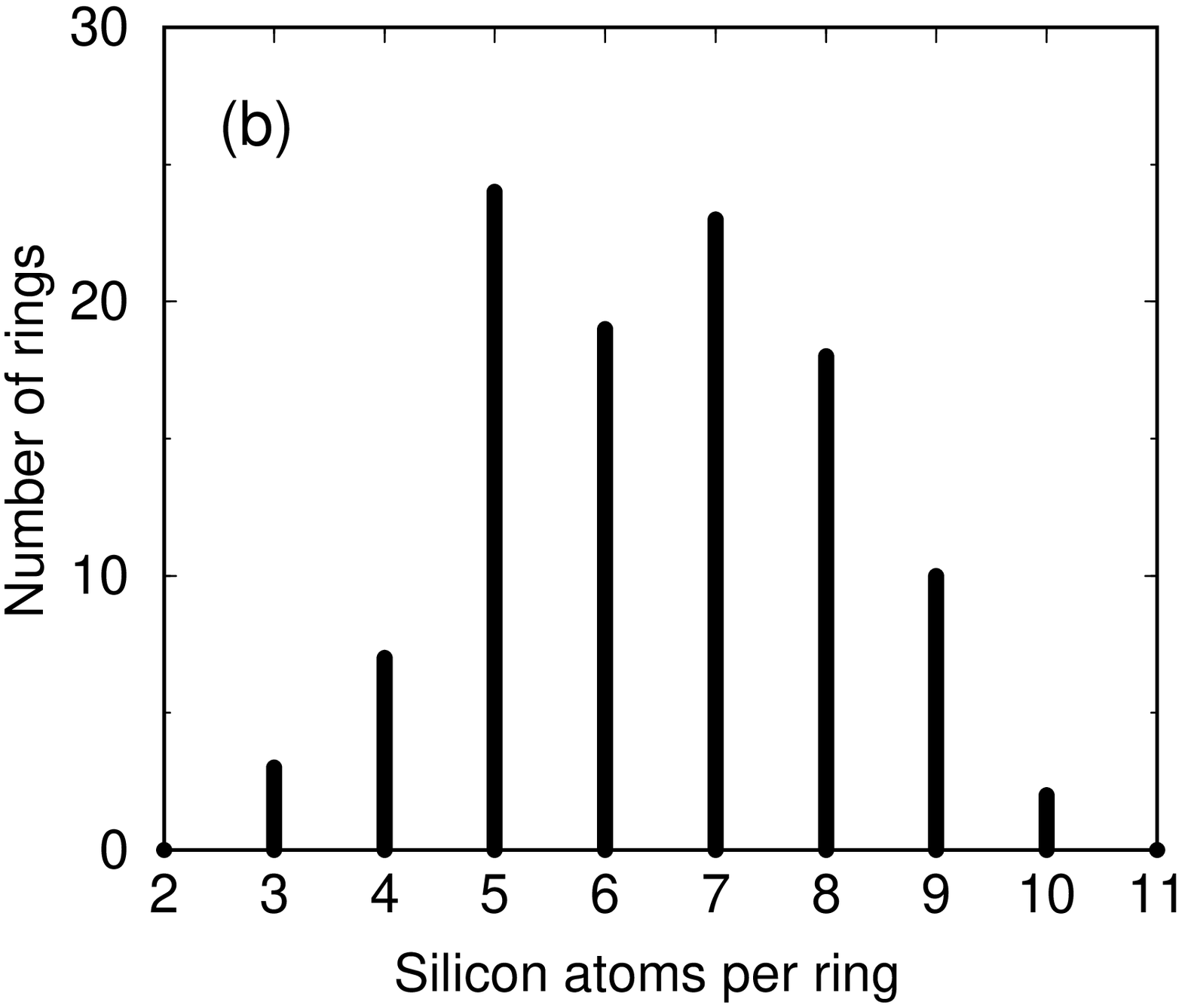}}
  \caption{The ring-size distributions of the 81-atoms (a) and of the 162-atoms silica models (b) 
           computed according to Ref. \cite{franz}.}
\end{center}
\label{rings}
\end{figure}

 The average O$\widehat{\rm Si}$O angle is very close to the
tetrahedral  value  (109.5$^o$), while the Si\^{O}Si bond angle distributions are quite different 
 in the classical and CPMD simulations.
The main effect
of  ab-initio annealing is a shift to lower angles of the
Si\^OSi angle distribution. The 
Si-O and O-O pair correlation functions are not  affected, while the Si-Si 
pair correlation function reflects the modification in the Si\^OSi angle distribution,
showing a longer tail at small Si-Si distances.
The ab-initio annealing and optimization does not affect the topology acquired by
the system during the classical MD quenching.
The ring-size distribution of the two a-SiO$_2$ models is reported in Fig. \ref{rings}.
The 81-atoms model produced with the low quenching rate has a low concentration of small
rings (3- and 4-membered) with respect to the 162-atoms model and to previous models
(72-atoms large) generated fully ab-initio \cite{sarnthein95}.
After the ab-initio annealing  performed with the softer Vanderbilt pseudopotentials,
the final structure 
has been further optimized  with norm-conserving pseudopotentials (for the
calculations of the dielectric properties).  
 The cell geometry has been optimized
at fixed volume  allowing orthorhombic distortions
of the initially cubic supercell such as to produce a diagonal  stress tensor.

\begin{figure}[!ht]
\centerline{\epsfxsize= 8.5 truecm\epsffile{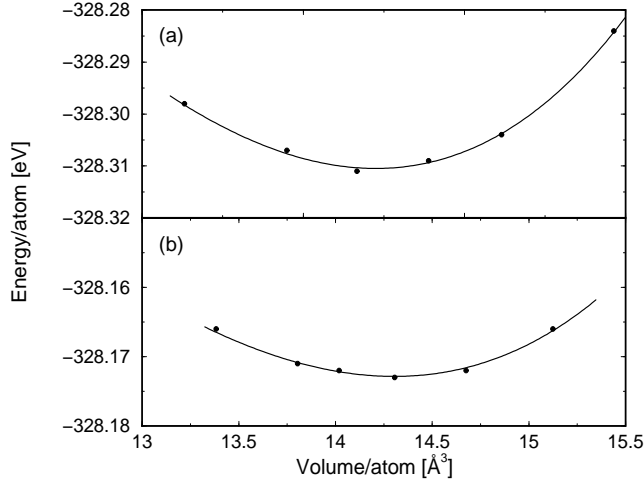}}
\caption{Ab-initio equation of state of the a-SiO$_2$ models.
a) 81-atoms supercell.  b) 162-atoms supercell.}
\label{EOS}
\end{figure}

The residual anisotropy in the stress ($\sigma$) is: 
\begin{equation}
\sigma=\left(\begin{array}{ccc}

          -422.8  &   -11.1 &    7.1 \\
           -11.1  &  -421.1 &   -2.6  \\
             7.1  &  -2.6   & -422.2 \\
         \end{array}  \right)\ \ \
\label{sigma81}
\end{equation}
for the optimized ratios $b/a$=1.051, $c/a$=1.002 in the 81-atoms supercell and
\begin{equation}
\sigma=\left(\begin{array}{ccc}
          -429.0  &  -14.0   &    8.4 \\
           -14.0  &  -424.5 &   6.1  \\
             8.4  &   6.1   & -424.3 \\
         \end{array}  \right)\ \ \
\label{sigma162}
\end{equation}
for the optimized $b/a$=1.037, $c/a$=1.018  in the 162-atoms supercell. The large negative stress 
 in eq. \ref{sigma81}
is  due to the so--called Pulay stress. 
The $b/a$ and $c/a$ ratio  obtained in this way at the initial density of 2.21 g/cm$^3$
have been then hold fixed
and the volume varied to generate the equation of state reported in Fig. \ref{EOS}.
As far as we know this is the first ab-initio calculation of the equation of state of
a-SiO$_2$.
The calculated $E(V)$ points have been corrected for the discontinuities
due to the incomplete basis set following the prescription given in ref. \cite{francis90} and
then fitted by a Murnaghan function \cite{murna}. The resulting equilibrium density ($\rho_{eq}$),
 bulk modulus ($B$) and derivate of the bulk modulus with respect to pressure ($B'$)
for the 162-atoms and 81-atoms models are $\rho_{eq}=2.34$ g/cm$^3$, $B=39.0$ GPa,
$B'=-7.01$ and $\rho_{eq}=2.31$ g/cm$^3$, $B=46.2$ GPa,
$B'=-6.28$, respectively. These results, especially for the
larger model, are in good agreement with the experimental values of
$\rho_{eq}=2.205$ g/cm$^3$, $B=36.7$ GPa,
$B'=-5.31$ \cite{kondo81}. 
Note in particular that the highly unusual negative value of $B'$ is well reproduced by our
calculations.
The bulk modulus would probably improve by
optimizing  $b/a$ and $c/a$ at  each volume,  held fixed in the equation of state of Fig. \ref{EOS}.
The error in $B$ is smaller for the larger system which has less geometrical constraints and can more
easily respond to the density change.
\begin{table}[!ht]
\begin{center}
\begin{tabular}{lccc}
\hline\hline
  size (atoms)           &  81   &   162  &   648  \\
\hline
  $\rho_{eq}$ (g/cm$^3$) & 2.23  & 2.30   &  2.29  \\
  B (GPa)                & 53.5  & 40.5   &  34.1  \\
  B$'$                   & -9.0  & -4.3   &  -6.3  \\
\hline\hline
\end{tabular}
\end{center}
\caption{ Equilibrium density ($\rho_{eq}$), bulk modulus ($B$) and first order derivative of
          the bulk modulus ($B'$) of amorphous silica models optimized with the BKS potential.}
\label{eos_cl}
\end{table}
In fact, we have also computed  the equation of state within
the classical molecular dynamics scheme, using the empirical
potential of ref. \cite{vanbeest} for the 81-, 162-atoms models and
for a larger amorphous model containing 648 atoms.
The results, reported in table \ref{eos_cl}, indicate that the
overestimation of the bulk modulus in the ab-initio calculations can be ascribed
mainly to a size effect.

Bond angle distributions of the fully optimized system at the equilibrium density are
plotted in figures \ref{struct81} and \ref{struct162} for the 81-atoms and for the 162-atoms 
model, respectively.
The shape  of the Si\^OSi angles distribution  strongly depends on the 
quenching rate. The model quenched at lower rate (81-atoms cell) has a 
larger mean Si\^OSi angle both in the classical simulation and after  ab-initio annealing.
Furthermore, for this model the concentration of small ($<$120$^0$) Si\^OSi
angles is lower than in the model generated fully ab-initio in ref. \cite{sarnthein95} and fits better to 
 experimental data.

\subsection{Dielectric and photoelastic properties}
Since the photoelastic coefficients are very sensitive to  density, we have computed the
dielectric properties of amorphous silica at the theoretical equilibrium volume (cfr. previous section).
The residual, small anisotropy in the structure  can also be identified
by inspection of the dielectric ($\varepsilon$) tensor:
\begin{equation}
\varepsilon=\left(\begin{array}{ccc}
          2.297   &  0.013  & -0.009 \\
          0.013   &  2.292  &  0.007 \\
         -0.009   &  0.007  &  2.288 \\
         \end{array}  \right)\ \ \ ,
\label{epsi81}
\end{equation}
for the 81-atoms supercell and
\begin{equation}
\varepsilon=\left(\begin{array}{ccc}
           2.192 &  0.009  &-0.004 \\
           0.009  &   2.198  &-0.018   \\
          -0.004  &   -0.018 & 2.191   \\
         \end{array}  \right)\ \ \
\label{epsi162}
\end{equation}
for the 162-atoms supercell.
For a fully isotropic and homogeneous system, as expected for
a truly amorphous system,  $\varepsilon$ should obviously reduce
to a single number.
The experimental value of $\varepsilon$ at $\rho=$2.2 g/cm$^3$ is
2.125 \cite{schroder80}.
At the same density the theoretical dielectric constant (1/3 Tr($\varepsilon$))
is $\varepsilon$=2.260 for the 81-atoms cell and $\varepsilon$=2.148 for the 162-atoms cell.
The misfit with respect to experiments is similar to those already
found for the crystalline phases of silica. For the largest system (162-atoms) 
the use of $\Gamma$ point only within the BP framework partially compensates
for the error due to LDA,
as occurs for $\beta$-cristobalite (cfr. section 3.5.3) resulting in an accidental better agreement
with experiments. The dielectric constant computed at $\Gamma$ within the BP for a 72-atoms cell of 
a-SiO$_2$ in ref. \cite{umari02} is 2.00, which confirms that,  within BP, $\varepsilon$
converges from below by increasing the cell size.

At the theoretical equilibrium density 
the dielectric constant 
is  $\varepsilon$=2.292 and $\varepsilon$=2.194 for the 81- and 162-atoms cells, respectively

We have  computed
the response to strain of the structural properties of the amorphous models 
as reported in
table \ref{asisiam}.
\begin{table}
\begin{center}
\begin{tabular}{lccc|cc}
\hline\hline
 &                               & 81-atoms                      &                               &  &   162-atoms  \\ 
 &  $\partial /\partial\eta_{11}$ &  $\partial /\partial\eta_{22}$ & $\partial /\partial\eta_{33}$ & $\partial /\partial\eta_{22}$ & $\partial /\partial\eta_{33}$\\
\hline
  $\overline{\rm{Si-O}}$    &  0.151    & 0.154  & 0.157 & 0.15   &  0.11 \\
  $\overline{\rm{Si-\widehat{O}-Si}}$ & 68.6    & 69.1   & 73.4&  68.5 & 57.5    \\
  P$_x$(Si-Si)         & 1.488   & -0.037    & -0.122 & -0.105  & -0.053   \\
  P$_y$(Si-Si)         & -0.141  &  1.553    & -0.049 &  1.452  & -0.044  \\
  P$_z$(Si-Si)         & -0.124  & -0.067    &  1.490 & -0.098  & 1.355   \\
\hline\hline
\end{tabular}
\end{center}
\caption{The derivatives of the SiO distance, SiOSi angles and  Si-Si vector distances 
         with respect to strains $\eta_{11}$, $\eta_{22}$ and $\eta_{33}$ in the models of a-SiO$_2$.
	 P$_i$(Si-Si) denotes the projection of the Si-Si vector distance on the $i$-th axis.}
	 
\label{asisiam}
\end{table}

The calculated photoelastic coefficients of the 81- and 162-atoms supercells
at the theoretical equilibrium density are reported
in table \ref{photoA}. 
The photoelastic coefficients have been calculated by finite differences
from the  dielectric tensor
of  systems with strains from -1 $\%$
to +1 $\%$.
As discussed in section II and IVb the calculation on the 81-atoms cell have been
performed within standard DFPT, while the calculations on the larger 162-atoms
supercell have been performed within the Berry phase.
Residual structural anisotropies are responsible for the differences among $p_{11}$,
$p_{22}$ and $p_{33}$ values and among $p_{12}$, $p_{21}$ and $p_{32}$.
A measure of the model isotropy is given by the fulfillment of the Cauchy relation
( 2$p_{44}$=$p_{11}-p_{12}$).
The value of $p_{44}$ 
computed by  applying the $\eta_{23}$ shear strain in the 81-atoms model 
is -0.074 (cfr. tab. \ref{photoA}) which
compares well with the value of
 ($p_{11}-p_{12})/2$, ranging from -0.069 to -0.092 which suggests a sufficient
 homogeneity of the small model as well.
\begin{table}[!ht]
\begin{center}
\begin{tabular}{lccc}
\hline\hline
           & 81-atoms  & 162-atoms &  exp. \\
\hline
$p_{11}$   &  0.072  &    -    & 0.125$^a$\ -\ 0.100$^b$ \\
$p_{22}$   &  0.045  &   0.047 & - \\
$p_{33}$   &  0.053  &   0.085 & - \\
$p_{12}$   &  0.217  &   0.223 & 0.27$^a$\ -\ 0.285$^b$ \\
$p_{21}$   &  0.230  &    -    & - \\
$p_{31}$   &  0.224  &    -    & - \\
$p_{32}$   &  0.209  &   0.227 & - \\
$p_{13}$   &  0.209  &   0.238 & - \\
$p_{23}$   &  0.197  &   0.235 & - \\
$p_{44}$   & -0.074  &     -   & -0.073$^a$ \\
\hline\hline
\end{tabular}
\end{center}
\caption{Photoelastic coefficients of the amorphous silica models containing 81 and 162
atoms compared to experimental data. $a$: ref. \cite{schroder80}, $b$: ref.\cite{vedam50}.}
\label{photoA}
\end{table}
Agreement with  experiments is fair, but less satisfactory than the performances 
of DFPT-LDA for crystalline phases. In particular, the ratio $p_{11}/p_{12}$ is largely 
underestimated in our model with respect to experiments. In view of the excellent results of
DFPT-LDA for crystalline silica, the discrepancy with experiments for a-SiO$_2$ could be naturally ascribed 
to an insufficient quality of our a-SiO$_2$ models, either due to a size effect or to a
still too fast quenching rate. 
However, we should also note that the measurement of the absolute values of $p_{11}$ and $p_{12}$
is very delicate and 
to our knowledgment only two independent measurements have been performed so far \cite{schroder80,vedam50}.

In order to obtain insight into the different microscopic contributions to photoelasticity and 
shed light on the relationship between structural properties and photoelastic response, we have 
developed a phenomenological model of photoelasticity in silica, as described in the next section.

\section{Phenomenological model of photoelasticity in silica}

In order to identify  the main
microscopic features ruling photoelasticity in silica,
we have devised a phenomenological model of the dielectric properties transferable
both to the crystalline and amorphous phases of SiO$_2$ 
by fitting on  the ab-initio database on photoelasticity in crystalline silica reported in the previous sections.

At this aim, we have assumed that the dielectric response of silica could be embodied in an ionic polarizability
tensor of the oxygen ions, whose value is assumed to depend on the Si\^OSi angle only. 
The dielectric tensor $\tens{\varepsilon}$ can be obtained  from a site dependent 
oxygen polarizability, $\tens{\alpha_i}$, as described below.
The dipole moment $\vec{p}_i$ at site $i$ is related to the 
local field $\vec{E}_i$ via the polarizability tensor $\tens{\alpha}$ as
\begin{equation}
  \vec{p}_i= \tens{\alpha_i} \vec{E}_i\ \ \ .
\label{defalpha}
\end{equation}
The local electric field $\vec{E}_i$ 
acting on a polarizable object at site $i$ is then \cite{kirkwood}:
\begin{equation}
  \vec{E}_i=\vec{D}-\sum^{N}_{k\neq i,\vec{R}} \tens{T^{\vec{R}}_{ik}}\cdot\vec{p}_k,
\label{localfield}
\end{equation}
where $\vec{D}$ is the electric displacement, $N$ the total number of polarizable centers per cell
and $\tens{T^{\vec{R}}_{ik}}$ is a symmetric tensor defined as:
\begin{equation}
  \tens{T^{\vec{R}}_{ik}} =\nabla_i\nabla_k\left(\frac{1}{r_{ik}}\right)=
  \frac{1}{r_{ik}^3}\left[1-3\frac{\vec{r}_{ik}\vec{r}_{ik}}{r_{ik}^2}\right]
\end{equation}
where $\vec{R}$ are the Bravais lattice vectors and $\vec{r}_{ik}$
is the distance between sites $i$ and $k$ in cells separated by  $\vec{R}$.
The sum of $\tens{T^{\vec{R}}_{ik}}$ over the Bravais lattice vectors $\vec{R}$ is performed as an
Ewald sum as described in ref. \cite{allentildsey}.

The electric displacement $\vec{D}$ is in turn  equal to 
$\vec{D}=\vec{E}+4\pi\vec{P}$, where $\vec{E}$ is  
 the macroscopic electric field $\vec{E}$ and 
 the electric polarization $\vec{P}$  can be expressed as
\begin{equation}
  {\vec P} = \frac{1}{V} \sum^N_i {\vec p}_i = \frac{1}{V} \sum^N_i \tens{\alpha_i} {\vec E}_i,
\label{pol_eloc}
\end{equation}
where  $V$ is the cell volume.
A set of linear equations  which relates 
the local field $\vec{E}_i$ to the macroscopic electric field is obtained
from eqs. \ref{localfield}-\ref{pol_eloc} as:
\begin{equation}
 \vec{E}=\vec{E}_i-\sum^{N}_{k\neq i}\left(\frac{4\pi}{V}-\sum_{\vec R}\tens{T^{\vec R}_{ik}}\right)\tens{\alpha_k}\vec{E}_k, 
 \label{mod1}
\end{equation}
or in a  compact matrix form
\begin{equation}
  \vec{E}=(\tens{I}-\tens{B})\lbrace\vec{E}_{i}\rbrace\ \ . 
\label{compact}
\end{equation}
$\tens{I}$ is the identity matrix, $\vec{E}$ and $\vec{E}_{i}$ are expressed as 3N vectors
and $\tens{B}$ is a 3Nx3N matrix consisting of 3x3 blocks $\tens{B_{ik}}$ defined as
\begin{equation}
  \tens{B_{ik}}=\left(\frac{4\pi}{V}-\sum_{\vec R}\tens{T^{\vec R}_{ik}}\right)\tens{\alpha_k}\ \ \ .
\end{equation}
For a  cubic lattice   
 equations \ref{mod1}
yield the Clausius-Mossotti (Lorentz-Lorenz) formula
(see ref. \cite{kirkwood} for details).
However, none of the crystalline polymorphs of SiO$_2$ is cubic, thus the term 
containing the dipole lattice sum ($\sum \tens{T_{ik}^{\vec{R}}}$) has to be explicitly evaluated.

From 
eq. \ref{pol_eloc} and eq. \ref{compact}, the following relation 
is obtained:
\begin{equation}
  \vec{P} = \frac{1}{V}\sum_{i,j} \tens{\alpha_i} \left(\tens{I}-\tens{B}\right)_{ij}^{-1}\vec{E} .
\label{polarization}
\end{equation}
The inversion of the matrix $\left(\tens{I}-\tens{B}\right)$ is performed in the 3N-dimensional space.
Finally, the dielectric susceptibility tensor $\tens{\chi}$
(${\vec P}=\tens{\chi}{\vec E}$)
results in:
\begin{equation}
  \tens{\chi} = \frac{1}{V}\sum_{i,j}^N \tens{\alpha_i} \left( \tens{I}-\tens{B} \right)_{ij}^{-1}.
\label{chi}
\end{equation} 
From $\tens{\chi}$, both the dielectric ($\tens{\varepsilon}=1+4\pi\tens{\chi}$) 
and the photoelastic tensors can be derived.

In order to exploit eq. \ref{chi}, in our empirical model we have assumed that the polarizable
centers in silica  are the oxygen ions. Secondly, we have also assumed 
that the oxygen polarizability depends on the Si\^OSi angle only.
These choices can be justified only a posteriori from comparison of the results produced 
by the phenomenological model with the ab--initio data. 
However, to support our model we can argue that most of the valence charge in SiO$_2$
is located on the oxygen ions and, secondly, that both the structural differences among the
different tetrahedral phases of silica (either crystalline or amorphous) and the
deformations induced by strain mainly rely on the distribution of the Si\^OSi angles.
The internal structure of the tetrahedra (O$\widehat{\rm Si}$O angles and Si-O bond lengths) 
is almost the same in the different polymorphs and undergoes minor deformations upon strain
since the open structure of silica allows to accommodate strain mostly by rotations of
the corner--sharing tetrahedra.

\begin{figure}[!ht]
\centerline{\epsfxsize= 7. truecm
\epsffile{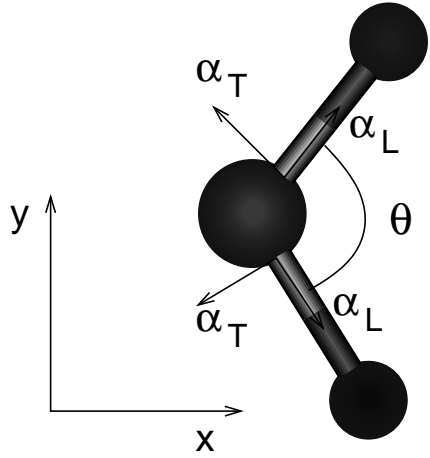}}
\caption{Sketch of  the Si-O-Si unit and of the bond contributions to polarizability.}
\label{modelfig}
\end{figure}
Our model of silica polarizability can also be understood as a particular case of the 
commonly used bond polarizability model (BPM). In the BPM the polarizability is expressed
as the sum of Si-O bond polarizabilities given by \cite{umari01}
\begin{equation}
  \alpha_{ij} = \frac{1}{3}(\alpha_L+2\alpha_T)\delta_{ij}+(\alpha_L-\alpha_T)
                \left(\frac{r_ir_j}{r^2}-\frac{1}{3}\delta_{ij}\right)
\label{bpmeq}
\end{equation}
where ${\vec r} = {\vec r}_{Si}-{\vec r}_O$ is the vector joining the silicon and oxygen atoms, and
$\alpha_L$ and $\alpha_T$ represent a longitudinal and a transversal polarizability of
the Si-O bond. $\alpha_L$ and $\alpha_T$ usually depend on 
the Si-O bondlength (see fig. \ref{modelfig}).
In our model we have introduced a second transverse bond polarizability,  $\alpha_{T'}$, 
in the direction orthogonal to the Si-O-Si plane (direction z in fig.  \ref{modelfig}).
We have made $\alpha_L$, $\alpha_T$ and $\alpha_{T'}$ dependent on the 
Si\^OSi angle ($\theta$), but independent on the Si-O bondlength. 
The dependence of the polarizability on the Si-O distance is crucial to reproduce 
Raman intensities, which are ruled by the modulation of the bond polarizability upon phonon 
displacements, that obviously involve both angle modulations and bond stretching.
Conversely, the change of $\alpha_L$, $\alpha_T$ and $\alpha_{T'}$ with Si-O bondlength
is not essential to describe photoelasticity, since strain-induced 
structural relaxations mainly consist of bond angles deformations not affecting the Si-O distances. 

\begin{figure}[!ht]
\centerline{\epsfxsize= 7. truecm
\epsffile{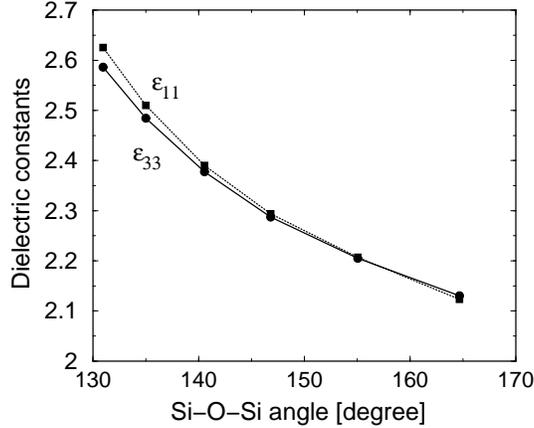}}
\caption{The dielectric tensor of $\alpha$-cristobalite as a function of the Si-$\widehat{\rm O}$-Si angle.}
\label{dielfit}
\end{figure}

\begin{figure}[!ht]
\centerline{\epsfxsize= 7. truecm
\epsffile{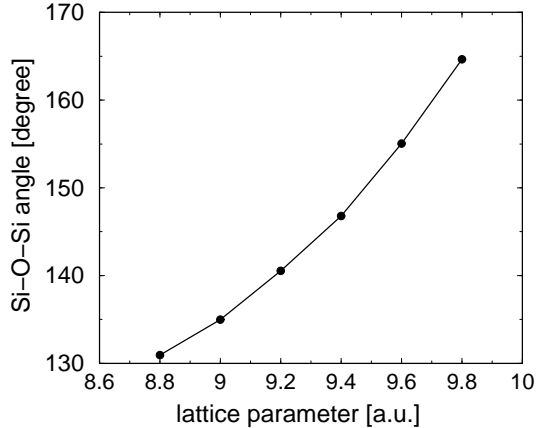}}
\caption{The modulation of the Si-$\widehat{\rm O}$-Si angle in $\alpha$-cristobalite as a function of 
         the lattice parameter $a$.}
\label{alphafit}
\end{figure}

In this approximation the dielectric response can be cast into a model based  on 
the polarizability of the oxygen ions only which is given by
\begin{equation}
 \tens{\alpha}= \left( \begin{array}{ccc}
   c(\theta)+\gamma cos^2(\theta /2) &   0  &  0 \\
   0  &    c(\theta)+\gamma sin^2(\theta /2) &  0 \\
   0 & 0 & \alpha_{T'}(\theta)
   \end{array} \right)
\label{polar}
\end{equation}
with $c=2(\alpha_L + \alpha_T + \alpha_{T'})/3$ and $\gamma=\alpha_L -\alpha_T$
(cfr. fig. \ref{modelfig} for axis orientations).
The contribution of each polarizable unit to the dielectric susceptibility is 
\begin{equation}
 \tens{\alpha_i} = \tens{R_i^T}\tens{\alpha}(\theta)\tens{R_i}
\end{equation}
where $\tens{R_i}$ is the rotation matrix that operates the transformation from 
the local reference system represented in figure
\ref{modelfig} to the absolute reference system of the solid, in which the $i$-th Si-O-Si unit is embedded.
The parameters $c$, $\gamma$ and $\alpha_{T'}$ have to be  determined and, in principle,
all of them depend on $\theta$. 
We have decided to fit these parameters on the dielectric tensor
of $\alpha$-cristobalite at several densities (see fig. \ref{dielfit}),
since  $\alpha$-cristobalite  responds to compression
mainly modifying the Si-$\widehat{\rm O}$-Si angles, maintaining the Si-O
bondlength unchanged \cite{mauri00}. 
We have checked that changing the density of  $\alpha$-cristobalite from 2.83 to
2.05 g/cm$^3$ (the theoretical equilibrium density is 2.54 g/cm$^3$) 
the Si-$\widehat{\rm O}$-Si angle changes from 130$^o$ to 165$^o$  with
a maximum variation of the Si-O bondlength of 0.006 \AA\ 
(see fig. \ref{alphafit}). 

\begin{figure}[!ht]
\centerline{\epsfxsize= 7. truecm
\epsffile{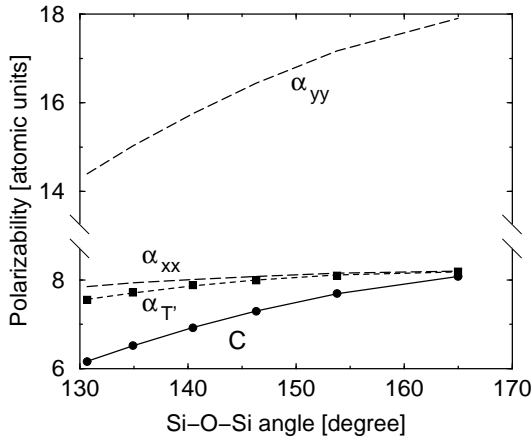}}
\caption{The functions $c(\theta)$ (solid line and circles) and $\alpha_{T'}(\theta)$ 
         (dashed line and squares), which assign the oxygen polarizability (see text). The data are the result 
         of the fitting  on the dielectric tensor of $\alpha$-cristobalite
         at different densities. The $\alpha_{xx}$ and $\alpha_{yy}$ components of the
         polarizability tensor (dashed lines) as obtained from eq. \ref{polar}.} 
\label{alphac}
\end{figure}

However, for $\alpha$-cristobalite the equations for $\chi$ (eq.
\ref{chi}) independent by symmetry are only two, 
not sufficient to fit all the three functions $c$, $\gamma$ and 
$\alpha_{T'}$ in eq. \ref{polar}.
For this reason, $\gamma$ has been assumed independent on $\theta$ and 
 set equal to the value  obtained from the fitting of the  ab-initio Raman spectrum of $\alpha$-quartz
within the BPM by Umari {\sl et al.} \cite{umari01} ($\gamma = 9.86$ a.u.).
The parameterization of the BPM proposed by Umari {\sl et al.}
reproduces the ab-initio Raman intensities of $\alpha$-quartz within  15$\%$.
The results for $c(\theta)$ and $\alpha_{T'}(\theta)$ obtained from 
 eq. \ref{chi} have been interpolated by  a second order polynomial 
(see fig. \ref{alphac}), whose
coefficients are reported in table \ref{fitpar}. 
\begin{table}[ht]
\begin{center}
\begin{tabular}{lccc}
\hline\hline
                        & a$_0$  &  a$_1$  &  a$_2$ \\
 \hline
  $c(\theta)$           & -20.46 &  18.38  & -2.942 \\
  $\alpha_{T'}(\theta)$ & -6.141 &  9.956  & -1.730 \\
\hline\hline
\end{tabular}
\caption{Coefficients of the polynomials that fit $c(\theta)$ and $\alpha_{T'}(\theta)$ ($\theta$ is
         expressed in radiants). }
\end{center}
\label{fitpar}
\end{table}

\begin{figure}[!ht]
\centerline{\epsfxsize= 7. truecm
\epsffile{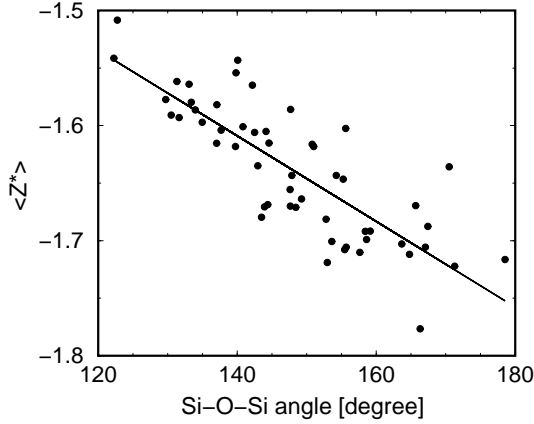}}
\caption{Dependence of the Born effective charges of oxygen ions on the Si-$\widehat{\rm O}$-Si
         angle for our 81-atoms model of a-SiO$_2$.}
\label{zeff}
\end{figure}

As shown in fig. \ref{alphac} the oxygen polarizability increases by increasing the Si-$\widehat{\rm O}$-Si
angle. In fact, the Si-O bond becomes more ionic by increasing the Si-$\widehat{\rm O}$-Si angle and
the larger charge on oxygen ion makes it more polarizable. 
The increased ionicity of Si-O bond at larger SiOSi angles is further confirmed by
the dependence of the effective charge of oxygen ions on the  SiOSi angles shown in \ref{zeff}.
The latter figure  reports the distribution of the effective charges ($1/3TrZ^*$) of oxygen ions 
in different sites of our  a-SiO$_2$ model, computed 
within DFPT \cite{baroni01}.

The transferability of the oxygen polarizability shown in fig. \ref{alphac} has been checked by 
comparing the dielectric constants of
the  other silica  polymorphs discussed in the previous sections, as 
obtained within the phenomenologial model and by DFPT (see table  \ref{modeldiel}).
The agreement is overall satisfactory.
In particular, birefringence of quartz and $\beta$-cristobalite are well reproduced.
\begin{table}
\label{modeldiel}
\begin{center}
\begin{tabular}{lccc}
\hline\hline
                   &                    & DFPT  &  model\\
\hline
$\alpha$-quartz    & $\varepsilon_{11}$ & 2.590 & 2.630 \\
P=0 GPa            & $\varepsilon_{33}$ & 2.620 & 2.670 \\
\hline
$\alpha$-quartz    & $\varepsilon_{11}$ & 2.694 & 2.775 \\
P=3 GPa            & $\varepsilon_{33}$ & 2.737 & 2.818 \\
\hline
$\alpha$-quartz    & $\varepsilon_{11}$ & 2.812 & 2.912 \\
P=7 GPa            & $\varepsilon_{33}$ & 2.865 & 2.970 \\
\hline
$\beta$-crist.     & $\varepsilon_{11}$ & 2.281 & 2.252 \\
                   & $\varepsilon_{33}$ & 2.251 & 2.227 \\
\hline
                   & $\varepsilon_{11}$ & 2.297 & 2.285 \\
a-SiO$_2$          & $\varepsilon_{22}$ & 2.292 & 2.281 \\
                   & $\varepsilon_{33}$ & 2.288 & 2.275 \\
\hline\hline
\end{tabular}
\caption{Comparison of the dielectric constants computed within  ab-initio DFPT
         and within the phenomenological model fitted on $\alpha$-cristobalite.}
\end{center}
\end{table}
The discrepancies for quartz at high pressure may be due to the change of the Si-O
bondlength upon compression (see table \ref{pression}), whose effect on the oxygen polarizability 
has been neglected in our model. This effect is less important in $\alpha$-cristobalite, whose 
more open structure can easily accommodate density changes with minor effect on the Si-O bondlength.
The phenomenological model reproduces  the dielectric constant of a-SiO$_2$ as well.

The transferability of the model has  been further checked by computing the photoelastic tensor.
This is given by the derivative of the 
dielectric susceptibility tensor $\chi$ (eq. \ref{chi}) with respect to the strain tensor 
$\eta_{\lambda\mu}$  as
\begin{eqnarray}
  \frac{\partial\tens{\chi}}{\partial\eta_{\lambda\mu}} = -\tens{\chi}\delta_{\lambda\mu} +  
      \frac{1}{V} \sum_i^N \Bigg(\left[ \tens{R_i^T}\frac{\partial\tens{\alpha}}{\partial\eta_{\lambda\mu}}\tens{R_i}\right]
      \tens{A} + \nonumber\\
     +\left[\tens{R_i^T}\tens{\alpha}\tens{R_i}\right]\frac{\partial\tens{A}}{\partial\eta_{\lambda\mu}}+
 \left[\frac{\partial \tens{R_i^T}}{\partial\eta_{\lambda\mu}} \tens{\alpha} \tens{R_i} 
 + \tens{R_i^T}\tens{\alpha}\frac{\partial \tens{R_i}}{\partial\eta_{\lambda\mu}}\right]\tens{A}\Bigg)
\label{dchi}
\end{eqnarray}
where the matrix $\tens{A}$ is defined as $\tens{A}=(\tens{I}-\tens{B})^{-1}$ (see eq. \ref{chi}), 
and the arguments in square brackets indicate
3x3N matrices. 
The change of the oxygen polarizability with strain can be expressed in terms of 
$\partial c(\theta)/\partial\theta$ and $\partial\alpha_{T'}(\theta)/\partial\theta$ deduced from Fig. \ref{alphac}.
All the other derivatives have been obtained by finite differences.
Eq. \ref{dchi} can provide an estimate of the photoelastic coefficients of all silica polymorphs 
made of corner--sharing tetrahedra.

The photoelastic coefficients calculated within the phenomenological model for a few silica
polymorphs are compared to ab-initio data in table \ref{photomod} (second and first column, respectively).
The same comments made above for the dielectric constants still hold for the photoelastic tensor. 
The agreement with ab--initio data is better for system at ambient conditions than at high pressure.
Overall we can conclude that the phenomenological model has good transferability. 

Table \ref{photomod} reports also the contribution from the individual terms 
supplying to the expression of $\partial\chi/\partial\eta_{\lambda\mu}$ in eq. \ref{dchi}.
\begin{table}
\begin{center}
\begin{tabular}{lcccccc}
\hline\hline
                   &          & DFPT  & model &  $\partial R/\partial\eta$ & Lorentz & $\partial\alpha/\partial\theta=0$ \\
\hline
$\alpha$-quartz    & $p_{13}$ & 0.240 & 0.301 &   0.043     & 0.278 & 0.469 \\
 P=0 GPa           & $p_{33}$ & 0.085 & 0.083 &  -0.087     & 0.067 & 0.245 \\
\hline

$\alpha$-quartz    & $p_{13}$ & 0.215 & 0.313 &   0.038     & 0.296 & 0.452 \\
 P=7 GPa           & $p_{33}$ & 0.067 & 0.079 &  -0.078     & 0.105 & 0.239 \\
\hline

                   & $p_{11}$ & 0.124 & 0.090 &  -0.082     & 0.049 & 0.290 \\
$\beta$-cristobalite     & $p_{21}$ & 0.291 & 0.275 &   0.040     & 0.236 & 0.470 \\
                   & $p_{31}$ & 0.269 & 0.236 &   0.041     & 0.207 & 0.351 \\
\hline
                   & $p_{11}$ & 0.072 & 0.103 &  -0.082     & 0.074 & 0.168 \\
a-SiO$_2$          & $p_{21}$ & 0.230 & 0.217 &   0.042     & 0.227 & 0.319\\
                   & $p_{31}$ & 0.224 & 0.214 &   0.038     & 0.221 & 0.306\\
\hline
                      & $p_{11}$ & 0.218 & 0.161 & -0.013   & 0.100 & 0.353 \\
                      & $p_{21}$ & 0.244 & 0.276 & -0.010   & 0.275 & 0.363 \\
$\alpha$-cristobalite & $p_{31}$ & 0.293 & 0.319 &  0.025   & 0.289 & 0.408 \\
                      & $p_{13}$ & 0.293 & 0.325 &  0.046   & 0.296 & 0.449 \\
                      & $p_{33}$ & 0.152 & 0.154 & -0.091   & 0.105 & 0.228 \\
\hline\hline
\end{tabular}
\end{center}
\caption{Comparison of the photoelastic coefficients yielded by the phenomenological model and by ab-initio
         DFPT. The data on a-SiO$_2$ refers to the 81-atoms model. See text for the description of the other columns.}
\label{photomod}
\end{table}

The first term in the right-hand side of eq. \ref{dchi} ($-\tens{\chi}\delta_{\lambda\mu}$) comes from 
the derivative of the density, which is null for strains preserving the volume i.e. for $\lambda\neq\mu$. 
The contribution of this term is always positive.
The second term in eq. \ref{dchi} contains the derivative of the polarizability of the Si-O-Si units. It can be expressed
as a function of the Si-$\widehat{\rm O}$-Si bond angle and depends on the derivative of the 
functions $c(\theta)$ and $\alpha_{T'}(\theta)$:
\begin{equation}
  \frac{\partial\alpha(\theta)}{\partial\eta}=\frac{\partial\alpha(\theta)}{\partial\theta}\frac{\partial\theta}{\partial\eta}
\end{equation}
with
\begin{equation}
   \frac{\partial\alpha(\theta)}{\partial\theta}=\left( \begin{array}{ccc}
    \frac{\partial c(\theta)}{\partial\theta}-0.5 \gamma sin(\theta) & 0 & 0 \\
    0 & \frac{\partial c(\theta)}{\partial\theta}+0.5 \gamma sin(\theta) & 0 \\
    0 & 0 & \frac{\partial\alpha_{T'}(\theta)}{\partial\theta} 
    \end{array} \right). 
\end{equation}
This term provides a negative contribution to all the components of the photoelastic tensor.
The last column in table \ref{photomod} reports the photoelastic coefficients 
obtained from eq. \ref{dchi}, but neglecting the contribution from $\partial\alpha/\partial\theta$.
By comparing the second and the last columns in table \ref{photomod}, it is clear that the 
change in polarizability with the Si\^OSi angle is essential
to correctly reproduce the ab-initio data. Although it has not been explicitly displayed in eq. \ref{dchi},
the derivative of $\alpha$ also contributes to the term  $\partial A/\partial\eta_{\lambda\mu}$. 

The last term in eq. \ref{dchi} depends on the derivative of the rotation matrices $R_i$
with respect to strain. It is related to the geometrical
effect of alignment of the Si-Si axis of the Si-O-Si unit (cfr. fig. \ref{modelfig}) 
along the axis 
of a tensile strain. This effect results in an increase of the off-diagonal components
and a reduction of the diagonal components of the photoelastic tensor. This effect  accounts for most of
the observed difference between diagonal  and off-diagonal photoelastic coefficients
($p_{11}$ and $p_{12}$ for instance).
This is proven by the results in the third column of table \ref{photomod}, which
reports the 
contribution to the photoelastic coefficients due  to the last 
term in eq. \ref{dchi} only. In fact the difference $p_{11}-p_{12}$ for a-SiO$_2$ and $\beta$-cristobalite,
and $p_{33}-p_{13}$ for $\alpha$-cristobalite and $\alpha$-quartz, in the second column of table \ref{photomod},
are close to the results in the third column. 
We note that the identification of the alignment of the Si-O-Si units as the source of a
large value of $p_{11}-p_{12}$ and $p_{33}-p_{13}$ is consistent with the structural
response data in tables \ref{psisiq}, \ref{asisiq} and \ref{asisiam}. 
The small value of  $p_{11}-p_{12}$ in $\alpha$-cristobalite is consistent with the weak
e alignment of the tetrahedra since 
for strain along the x-axis is weak (cfr. table \ref{asisiq}).

The term depending on the derivative of the matrix $\tens{A}$ accounts for the 
the presence of local fields correction  and its modification upon strain.
Its contribution to the photoelastic coefficients is inferred by computing 
 $p_{ij}$ within the Lorenz-Lorentz approximation. 
The results, shown in the 4th column of tab. \ref{photomod}, demonstrate that 
local field effects beyond the Lorenz-Lorent approximation (i.e. inclusion of anisotropic local
fields produced by near molecules)
modify the photoelastic coefficients for crystalline phases
up to 45$\%$.
As expected, corrections to the Lorenz-Lorentz approximation are small for
the amorphous silica models.
The phenomenological model allowed us to address the role of the size of the simulation cell on
the photoelastic tensor of a-SiO$_2$. This has been done by computing the photoelastic coefficient
within the phenomenological model for the same 81-atoms and 162-atoms models discussed so far and for a larger
model of 648 atoms, all optimized with the BKS potential.  The results, reported in table \ref{photoclass},
quantify the errors due to finite size effects. Isotropy in the diagonal components $p_{ii}$ are achieved only 
for the largest model. However, errors due to finite size effects for cells containing 81-162 atoms are 
only slightly larger than the errors due to LDA-DFT.

\begin{table}[!ht]
\begin{center}
\begin{tabular}{lccc}
\hline\hline
	   & 81-atoms  & 162-atoms &  648-atoms \\
	   \hline
           $p_{11}$   & 0.115   & 0.100  & 0.132 \\
           $p_{22}$   & 0.161   & 0.143  & 0.141 \\
           $p_{21}$   & 0.227   & 0.231  & 0.234 \\
           $p_{31}$   & 0.237   & 0.237  & 0.227 \\
           $p_{12}$   & 0.245   & 0.216  & 0.220 \\
           $p_{32}$   & 0.240   & 0.239  & 0.220 \\
\hline\hline
	   \end{tabular}
	   \end{center}
	   \caption{Photoelastic coefficients computed within the phenomenological model for
	   three models of a-SiO$_2$ optimized with the BKS classical potential.}
	   \label{photoclass}
	   \end{table}

\section{Discussion and Conclusions}

Photoelasticity of crystalline and amorphous silica has been 
studied by  density functional perturbation theory. 
This framework 
has been  checked successfully first on paradigmatic semiconducting and
insulating systems, silicon and MgO.
The same framework has been applied to $\alpha$-quartz providing photoelastic constants in good agreement
with previous calculations \cite{detraux01} and with experiments.
 The photoelastic coefficients
have been computed also for $\alpha$-quartz under pressure up to 7 GPa, and for 
$\alpha$-cristobalite and $\beta$-cristobalite at normal conditions.
The theoretical results obtained for these latter systems
are predictions open to experimental verification. 

We have thus verified that the simple DFPT-LDA framework is suitable to reproduce accurately 
the photoelastic coefficients. Correction of the  DFT-LDA band gap by a scissor  operator 
 is necessary to reproduce the dielectric constants but has minor effects
on the photoelastic tensor. 

We have then studied photoelasticity for models of amorphous silica, 
containing up to 162 atoms, generated by quenching from the
melt in molecular dynamics simulations. The calculated photoelastic coefficients 
are in good agreement with experimental data. Our ab-initio framework has thus valuable 
predictive power also for the amorphous phase. 

Along with the calculated photoelastic coefficients, we have collected data on the 
structural response to strain of different crystalline phases. The analysis of these data 
has aided us in the development of a phenomenological model of photoelasticity in silica
polymorphs. In fact the analysis of the structural changes upon strain has shown that silica
responds to strain mainly via rotations of the tetrahedral units (thus changing the
Si-\^O-Si angles) with smaller changes of the intra-tetrahedral 
(O-$\widehat{\rm Si}$-O) angles. For instance, the most prominent effect of tensile stress is
the alignment of the Si-O-Si units to the direction parallel to the strain axis.

These results suggested us to model the dielectric properties of silica polymorphs in terms of 
Si-O-Si polarizable units with polarizability dependent on the Si-\^O-Si angle only. 
At this aim we have developed a phenomenological model fitted on the dielectric constant of
$\alpha$-cristobalite at different densities.
The model consists of polarizable oxygen ions with anisotropic polarizability 
dependent on the Si-\^O-Si angle. 
The validity of the model has been proven by comparing dielectric
and photoelastic coefficients computed within the phenomenological model to the results 
of DFPT calculations.
The agreement is good and the simple phenomenological model sheds light on the main contributions
to photoelasticity in silica polymorphs.
It turns out that the change in the oxygen polarizability with strain, via the Si\^OSi angle is 
essential to reproduce the photoelastic response. The neglect of this term introduces errors of up to
 50 $\%$. Secondly, the difference in the diagonal and off-diagonal photoelastic coefficients
(e.g. $p_{11}$ and $p_{21}$), which is particularly large in a-SiO$_2$, is due to the alignment of the
SiOSi units along the axis of tensile strain. 
Finally, while local fields beyond the Lorentz-Lorent approximation (via dipole-dipole interactions) 
 and their change upon to strain are important to reproduce the dielectric and photoelastic 
properties (expecially birefringence) of the low-symmetry crystalline phases, 
 they are obviously negligible for a-SiO$_2$.
This result confirms that our models of a-SiO$_2$, despite their small size, are already good model of
a homogeneous isotropic amorphous network.

In summary, our work has demonstrated the reliability of the ab-initio DFT-LDA framework in the
study of photoelasticity of silica systems and  has provided a phenomenological model of the dielectric properties
transferable to several crystalline and amorphous polymorphs.
The same ab-initio framework, supplemented by the phenomelogial model, could also provide a microscopic description of
the photoelasic properties of more complex silica-based glasses.

\section{Acknowledgments}

D.D. acknowledges Pirelli Cavi e Sistemi S.p.a. for financial support.
This work has been partially supported by the INFM Parallel Computing Initiative.

\end{document}